   \def\versionno{ n=2starsound -- draft   }

\catcode`\@=11

\expandafter\ifx\csname draftcontrol\endcsname\relax\global\def\draftcontrol{0}
\fi

{\count255=\time\divide\count255 by 60
\xdef\hourmin{\number\count255}
\multiply\count255 by-60\advance\count255 by\time
\xdef\hourmin{\hourmin:\ifnum\count255<10 0\fi\the\count255}}
\def\draftdate{\number\month/\number\day/\number\year\ \ \ \hourmin }

\newcommand\makepapertitle{\par
  \begingroup
    \renewcommand\thefootnote{\@fnsymbol\c@footnote}%
    \def\@makefnmark{\rlap{\@textsuperscript{\normalfont\@thefnmark}}}%
    \long\def\@makefntext##1{\parindent 1em\noindent
            \hb@xt@1.8em{%
                \hss\@textsuperscript{\normalfont\@thefnmark}}##1}%
     \newpage
     \global\@topnum\z@   
     \@makepapertitle
     \thispagestyle{empty}\@thanks
  \endgroup
  \setcounter{footnote}{0}%
  \global\let\thanks\relax
  \global\let\makepapertitle\relax
  \global\let\@makepapertitle\relax
  \global\let\@thanks\@empty
  \global\let\@author\@empty
  \global\let\@date\@empty
  \global\let\@title\@empty
  \global\let\title\relax
  \global\let\author\relax
  \global\let\date\relax
  \global\let\and\relax
  \def\version{\let\version\@version\@gobble}
}
\def\@makepapertitle{%
  \newpage
   \ifnum\draftcontrol=1 {}
   \version\versionno
   \vskip 3em%
   \else
   \hfill\hbox to 3cm {\parbox{4cm}{\@pubnum}\hss}%
   \vskip 3em%
   \fi
   \begin{center}%
   \let \footnote \thanks
     {\LARGE {\@title}}%
     \vskip 1.5em%
     {\normalsize
       \lineskip .5em%
       \begin{tabular}[t]{c}%
         \@author
       \end{tabular}\par}%
     \vskip 1.5em%
     {\@bstract}%
     \end{center}%
     \vskip 1.5em
     \@date%
   \par
}

\gdef\@pubnum{}
\def\pubnum#1{%
  \gdef\@pubnum{#1}}

\gdef\@bstract{}
\def\Abstract#1{%
  \gdef\@bstract{%
   \parbox{\textwidth-0pc}{%
   \centerline{\bf Abstract}\penalty1000%
\kern.2cm%
\noindent
\renewcommand\baselinestretch{1.0}%
{#1}}}
}

\def\ps@paper{\let\@mkboth\@gobbletwo%
     \ifnum\draftcontrol=1
    \def\@oddfoot{\hbox to \textwidth{\tiny \versionno \hfil\tiny\draftdate}%
    \hskip -\textwidth \hbox to \textwidth{\hfil\rm\thepage\hfil}}%
     \else\def\@oddfoot{\hbox to \textwidth{\hfil\rm\thepage\hfil}}
     \fi
     \let\@evenfoot\@oddfoot
}

\def\body{\clearpage
          \pagestyle{paper}
    }

\def\@version#1{\ifnum\draftcontrol=1
\typeout{}\typeout{#1}\typeout{}
\vskip3mm\centerline{\hbox{\fbox{\normalsize{\tt DRAFT -- #1 -- }
                   {\draftdate}}}}\vskip3mm
\fi}
\let\version\@version
\long\def\eqlabel#1{\ifnum\draftcontrol=1
                    \tag@false  
                    \tag*{(\theequation) \hbox to -0.2cm{\hspace{0cm}\small{#1}\hss}}
                    \refstepcounter{equation}
                    \edef\@currentlabel{\theequation}
                    \ltx@label{#1}          
                    \else
                    \label{#1}
                    \fi
                    }
\let\st@bibitem\@bibitem
\let\st@lbibitem\@lbibitem
\ifnum\draftcontrol=1
  \def\@bibitem#1{%
    \st@bibitem{#1}\a@@label{#1}\ignorespaces}
  \def\@lbibitem[#1]#2{%
    \st@lbibitem[#1]{#2}\a@@label{#2}\ignorespaces}
  \def\a@@label#1{%
    \gdef\a@lab{\smash{\normalfont\small#1}}
    \ifvmode
      \if@inlabel
        \global\setbox\@labels\hbox{%
          \llap{\a@lab\let\a@lab\relax
                \kern\@totalleftmargin\kern\marginparsep}%
          \box\@labels}%
      \fi
    \fi}
\fi

\documentclass[12pt,letterpaper]{article}

\usepackage{amsmath,amssymb,array,calc,epsfig}

\ifnum\draftcontrol=1
\fi

\renewcommand\baselinestretch{1.25}
\renewcommand\section{\@startsection {section}{1}{\z@}%
                                   {-3.5ex \@plus -1ex \@minus -.2ex}%
                                   {2.3ex \@plus.2ex}%
                                   {\normalfont\large\bfseries}}
\renewcommand\subsection{\@startsection{subsection}{2}{\z@}%
                                   {-3.25ex\@plus -1ex \@minus -.2ex}%
                                   {1.5ex \@plus .2ex}%
                                   {\normalfont\normalsize\bfseries}}
\renewcommand\subsubsection{\@startsection{subsubsection}{3}{\z@}%
                                   {-3.25ex\@plus -1ex \@minus -.2ex}%
                                   {1.5ex \@plus .2ex}%
                                   {\normalfont\normalsize\it}}
\renewcommand\paragraph{\@startsection{paragraph}{4}{\z@}%
                                   {-3.25ex\@plus -1ex \@minus -.2ex}%
                                   {1.5ex \@plus .2ex}%
                                   {\normalfont\normalsize\bf}}

\numberwithin{equation}{section}

\def\ie{{\it i.e.}}

\def\revise#1       {\raisebox{-0em}{\rule{3pt}{1em}}%
                     \marginpar{\raisebox{.5em}{\vrule width3pt\
                     \vrule width0pt height 0pt depth0.5em
                     \hbox to 0cm{\hspace{0cm}{%
                     \parbox[t]{4em}{\raggedright\footnotesize{#1}}}\hss}}}}

\newcommand\nxt[1]  {\\\fnxt#1}

\def\calc         {{\cal C}}

\def\cale         {{\cal E}}
\def\calf         {{\cal F}}

\def\cali         {{\cal I}}

\def\call         {{\cal L}}
\def\calm         {{\cal M}}
\def\caln         {{\cal N}}
\def\calo         {{\cal O}}
\def\calp         {{\cal P}}

\def\del          {\partial}

\def\ee           {{\rm e}}

\def\tr           {\mathop{\rm Tr}}

\def\sqr#1#2{{\vcenter{\vbox{\hrule height.#2pt
 \hbox{\vrule width.#2pt height#1pt \kern#1pt
 \vrule width.#2pt}\hrule height.#2pt}}}}

\newcommand{\fft}[2]{{\frac{#1}{#2}}}
\newcommand{\ft}[2]{{\textstyle{\frac{#1}{#2}}}}
\def\jsquare{\mathop{\mathchoice{\sqr{8}{32}}{\sqr{8}{32}}
{\sqr{6.3}{9}}{\sqr{4.5}{9}}}}

\def\a{\alpha}
\def\b{\beta}
\def\w{\omega}
\def\r{\rho}
\def\dd{\delta}

\def\c{\chi}
\def\ga{\gamma}
\def\ee{\cale}

\def\rh{\hat{\rho}}
\def\chih{\hat{\chi}}

\def\aa1{\phi}
\def\cc1{\psi}

\def\nm{\nabla_\mu}
\def\nn{\nabla_\nu}
\def\G{\Gamma}
\def\arctanh{{\rm arctanh}}

\catcode`\@=12

\begin{document}


\title{Sound waves in strongly coupled non-conformal gauge theory plasma}

\pubnum{%
hep-th/0507026}
\date{July 2005}

\author{
Paolo Benincasa$ ^1$,  Alex Buchel$ ^{1,2}$ and Andrei O. Starinets$ ^2$\\[0.4cm]
\it $ ^1$Department of Applied Mathematics\\
\it University of Western Ontario\\
\it London, Ontario N6A 5B7, Canada\\[0.2cm]
\it $ ^2$Perimeter Institute for Theoretical Physics\\
\it Waterloo, Ontario N2J 2W9, Canada\\
}

\Abstract{Using gauge theory/gravity duality we study sound wave
propagation in strongly coupled non-conformal gauge theory plasma.
We compute the speed of sound and the bulk viscosity of
$\caln=2^*$ supersymmetric $SU(N_c)$ Yang-Mills plasma at a
temperature much larger than the mass scale of the theory in the
limit of large $N_c$ and large 't Hooft coupling. The speed of
sound is computed both from the equation of state and the
hydrodynamic pole in the stress-energy tensor two-point
correlation function. Both computations lead to the same result.
Bulk viscosity is determined by computing the attenuation constant
of the sound wave mode.}

\makepapertitle

\body

\version\versionno

\section{Introduction and summary}
The conjectured duality between gauge theory and string theory
 \cite{m9711,Gubser:1998bc,Witten:1998qj,Aharony:1999ti} is
 useful for insights into the dynamics of gauge
theories. For strongly coupled finite-temperature gauge theories
in particular, the duality provides an effective description in
terms of a supergravity background involving black holes or black
branes. A prescription for computing the Minkowski space
correlation functions in the gauge/gravity correspondence
\cite{ss,hs} makes it possible to study the real-time
near-equilibrium processes (e.g. diffusion and sound propagation)
in thermal gauge theories at strong coupling
\cite{ne1,ne2,ne3,ne4,ne5,bh1,kss1,bls,bh2}. Computation of
transport coefficients in gauge theories whose gravity or string
theory duals are currently known may shed light on the true values
of those coefficients in QCD, and be of interest for hydrodynamic
models used in theoretical interpretation of elliptic flows
measured in heavy ion collision experiments at RHIC
\cite{r1,r2,r3}. This expectation is supported by an intriguing
observation \cite{kss1, bh2, kss, bl1} that the ratio of the shear
viscosity to the entropy density in a strongly coupled gauge
theory plasma in the regime described by a gravity dual is
universal for all such theories and equal to $1/4\pi$. (Finite 't
Hooft coupling corrections to the shear viscosity of $\caln=4$
supersymmetric $SU(N_c)$ gauge theory plasma in the limit of
infinite $N_c$ were computed in \cite{bls}.)

Previously, it was shown \cite{ne4} that the dual supergravity
computations reproduce the expected dispersion relation for sound
waves in the strongly coupled $\caln=4$ supersymmetric Yang-Mills
(SYM) theory plasma
\begin{equation}
\w(q)=v_s q -i\ \frac{\Gamma}{2}\, q^2 + O(q^3)\,,
\eqlabel{dispertion}
\end{equation}
where
\begin{equation}
v_s = \left( \frac{\partial P}{\partial \ee}\right)^{1/2}
\end{equation}
is the speed of sound, $P$ is pressure and $\ee$ is the volume
energy density. The attenuation constant $\Gamma$ depends on shear
and bulk viscosities $\eta$ and $\zeta$,
\begin{equation}
\Gamma =\frac{1}{\ee +P} \left( \zeta + \frac{4}{3} \eta\right)\,.
\end{equation}
For homogeneous systems with zero chemical potential $\ee + P = s
\ T$, where $s$ is the volume entropy density.
 Conformal symmetry of the
$\caln=4$ gauge theory  ensures that
\begin{equation}
v_s= \frac{1}{\sqrt{3}},\qquad \zeta=0 \eqlabel{n4dis}\,.
\end{equation}
Indeed, in conformal theories the trace of the stress-energy
tensor vanishes, implying the relation between $\ee$ and $P$ of
the form $\ee=3P$, from which the speed of sound follows.

 Non-conformal gauge theories\footnote{One should add at once that
throughout the paper terms "conformal" and "non-conformal" refer
to the corresponding property of a theory at {\it zero
temperature}. Conformal invariance of $\caln=4$ SYM is obviously
broken at finite temperature, but we still refer to it as
"conformal theory" meaning that the only scale in the theory is
the temperature itself.}, and QCD in particular, are expected to
have non-vanishing bulk viscosity and the speed of sound different
from the one given in Eq.~\eqref{n4dis}. Lattice QCD results for
the equation of state $\ee = P (\ee)$ suggest that $v_s \approx
1/\sqrt{3}\approx 0.577$ for $T\simeq 2 T_c$, where $T_c \approx
173$ MeV  is the temperature of the deconfining phase transition
in QCD. When $T\rightarrow T_c$ from above, the speed of sound
decreases rather sharply (see Fig.~11 in \cite{Karsch:2003jg},
\cite{Gavai:2004se}, \cite{Heinz:2005ja},
and references therein). 
To the best of our knowledge, no results were previously available
for the bulk viscosity of non-conformal four-dimensional gauge
theories\footnote{Perturbative results for the bulk viscosity and
other kinetic coefficients in thermal quantum field theory of a
scalar field were reported in \cite{Jeon:1995zm}. Bulk viscosity
of Little String Theory at Hagedorn temperature was recently
computed in \cite{Parnachev:2005hh} using string (gravity) dual
description. The speed of sound of a four-dimensional (non-conformal) 
cascading gauge theory was computed in \cite{aby}, also from the 
dual gravitational description. 
}.

In this paper we take a next step toward understanding transport
phenomena in four-dimensional gauge theories at strong coupling.
Specifically, we study sound wave propagation in mass-deformed
$\caln=4$  $SU(N_c)$ gauge theory plasma. In the language of
four-dimensional $\caln=1$ supersymmetry, $\caln=4$ gauge theory
contains a vector multiplet $V$ and three chiral multiplets
$\Phi,Q,\tilde{Q}$, all in the adjoint representation of the gauge
group. Consider the mass deformation of the $\caln=4$ theory,
where all bosonic components of the chiral multiples $Q,
\tilde{Q}$ receive the same mass $m_b$, and all their fermion
components receive the same mass $m_f$. Generically, $m_f\ne m_b$,
and the supersymmetry is completely broken. When $m_b=m_f=m$ (and
at zero temperature), the mass-deformed theory enjoys the enhanced
$\caln=2$ supersymmetry with $V, \Phi$ forming an $\caln=2$ vector
multiplet and $Q,\tilde{Q}$  forming  a hypermultiplet. In this
case, besides the usual gauge-invariant kinetic terms for these
fields\footnote{The classical K\"ahler potential is normalized
according to $2/g_{YM}^2\tr [\bar{\Phi}\Phi+\bar{Q}Q+\bar{\tilde
Q}\tilde{Q}]$.}, the theory has additional interaction and
hypermultiplet mass terms given by the superpotential
\begin{equation}
W=\frac{2\sqrt{2}}{g_{YM}^2}\ \tr\left([Q,\tilde{Q}]\Phi\right)
+\frac{m}{g_{YM}^2}\left(\tr Q^2+\tr \tilde{Q}^2\right)\,.
\eqlabel{w}
\end{equation}
The mass deformation of $\caln=4$ Yang-Mills theory described
above  is known as the $\caln=2^*$ gauge theory. This theory has
an exact nonperturbative solution found in \cite{dw}. Moreover, in
the regime of a large 't Hooft coupling, $g_{YM}^2 N_c\gg 1$, the
theory has an explicit  supergravity dual description known as the
Pilch-Warner (PW) flow \cite{pw}. Thus this model provides an
explicit example of the gauge theory/gravity duality where some
aspects of the non-conformal dynamics at strong coupling can be
quantitatively understood on both gauge theory and gravity sides,
and compared. (The results were found to agree \cite{bpp,ejp}.)
Given that the $\caln=2^*$ gauge theory is non-conformal, and at
strong coupling has a well-understood dual supergravity
representation, it appears that its finite-temperature version
should be a good laboratory for studying transport coefficients in
non-conformal gauge theories.

The supergravity dual to finite-temperature strongly coupled
$\caln=2^*$ gauge theory was considered in \cite{bl}.
It was shown that the singularity-free nonextremal deformation of the 
full ten-dimensional supergravity background of Pilch and Warner \cite{pw} 
allows for a consistent Kaluza-Klein reduction to 
five dimensions.
These nonextremal geometries are characterized by three independent parameters, 
\ie, the temperature and the coefficients of the non-normalizable modes of two scalar fields $\alpha$ and 
$\chi$ (see Section 2 for more details). 
Using the standard gauge/gravity correspondence, from the asymptotic behavior of the scalars 
the coefficient of the non-normalizable mode of $\alpha$-scalar was identified with the coefficient 
of the mass dimension-two operator in the 
$\caln=4$ supersymmetric Yang-Mills theory, \ie, the bosonic mass term $m_b$;  
the corresponding coefficient of the 
$\chi$-scalar is identified with the fermionic mass term $m_f$. 
The supersymmetric Pilch-Warner flow \cite{pw} constraints those coefficients 
so that $m_b=m_f$. 
However, as emphasized in \cite{bl}, on the supergravity side $m_b$ and $m_f$ 
are independent, and for $m_b\ne m_f$ correspond to the mass deformation of $\caln=4$ 
which completely breaks supersymmetry. 
In the high temperature limits $m_b/T \ll 1$ (with $m_f\ll m_b$)
 or  $m_f/T\ll 1$ (with $m_b\ll m_f$),
the background black brane geometry was constructed
analytically.
 (For generic values of $m_b$, $m_f$, the background
supergravity fields satisfy a certain system of coupled nonlinear
ODEs \cite{bl}, which in principle can be solved numerically.) The
holographic renormalization of this theory was discussed in
\cite{bh1}, where it was also explicitly checked that for
arbitrary values of $m_b$, $m_f$ in the limit of large $N_c$ and
large 't Hooft coupling
 the ratio of the shear viscosity to the entropy density in
 the theory
equals $1/4\pi$, in agreement with universality
\cite{kss1,bh2,kss,bl1}.

Our goal in this paper is to compute the parameters of the
dispersion relation \eqref{dispertion} for the $\caln=2^*$ plasma.
Since the ratio $\eta/s$ is known, this will allow us to determine
the speed of sound and the ratio of bulk to shear viscosity in the
$\caln=2^*$ theory. We will be able to do it only in the
high-temperature regime where the metric of the gravity dual is
known analytically, leaving the investigation of the full
parameter space to future work.
The dispersion relation \eqref{dispertion} appears as a pole of
the thermal two-point function of certain components of the
stress-energy tensor in the hydrodynamic approximation, i.e. in
the regime where energy and momentum are small in comparison with
the inverse thermal wavelength ($\omega/T\ll 1$, $q/T \ll 1$).
Equivalently, Eq.~\eqref{dispertion} can be computed as the
 lowest (hydrodynamic) quasinormal frequency of a certain class of
gravitational perturbations of the dual supergravity background
\cite{set}.

Though the  general framework for studying sound wave propagation
in strongly coupled gauge theory plasma from the supergravity
perspective is known \cite{ne4,set}, the application of this
procedure to a non-conformal gauge theory is technically quite
challenging.
 The main difficulty stems from the fact that
unlike the (dual) shear mode graviton fluctuations, the (dual)
sound wave graviton fluctuations do not decouple from supergravity
matter fluctuations.
The interactions between various fluctuations and their background
coupling appear to be gauge-theory specific. As a result, we do
not expect speed of sound or bulk viscosity to exhibit any
universality property similar to the one displayed by the shear
viscosity in the supergravity approximation. In fact, we find that
the speed of sound and its attenuation do depend on the mass
parameters of the $\caln=2^*$ gauge theory.

Let us summarize our results.
 For the speed of sound and the ratio of the shear to bulk
viscosity we find, respectively,
\begin{equation}
v_s= \frac{1}{\sqrt{3}} \left( 1-\frac{ \left[\Gamma\left(\ft
34\right)\right]^4}{3\pi^4}\
 \left(\frac{m_f}{T}\right)^2
-\frac{1}{18\pi^4}\ \left(\frac{m_b}{T}\right)^4 +\cdots\right)\,,
\eqlabel{csa}
\end{equation}
\begin{equation}
\frac{\zeta}{\eta} = \beta_f^\Gamma \, \frac{
\left[\Gamma\left(\ft 34\right)\right]^4}{3\pi^3}\
 \left(\frac{m_f}{T}\right)^2 +  \frac{\beta_b^\Gamma}{432\pi^2}\
\left(\frac{m_b}{T}\right)^4 +\cdots\,, \eqlabel{csax}
\end{equation}
where $\beta_f^\Gamma \approx 0.9672$,
 $\beta_b^\Gamma \approx 8.001$, and
 the ellipses
 denote higher order terms in $m_f/T$ and $m_b/T$.
From the dependence \eqref{csa}, \eqref{csax} it follows that at
least
 in the high temperature regime the
ratio of bulk viscosity to shear viscosity is proportional to the
deviation of the speed of sound squared from its value in
conformal theory,
\begin{equation}
\frac{\zeta}{\eta} \simeq - \kappa \, \left(
v_s^2-\frac{1}{3}\right)\,,
\eqlabel{proporo}
\end{equation}
where\footnote{In general 
$\kappa=\kappa(\lambda)$ is a function of the ratio $\lambda\equiv 
m_b/m_f$ which we were able to compute in two limits $\lambda\to 0$
 and $\lambda\to \infty$. 
Assuming $\kappa(\lambda)$  to be a smooth monotonic function,
 we find that it  varies 
by $\sim 8\%$ over the whole range  $\lambda\in [0,+\infty)$. 
Additionally, 
for both finite (and small) $m_b/T,\ m_f/T$ we verified scaling
 \eqref{proporo} by
 explicit fits.}
  $\kappa=3\pi\beta_f^\Gamma/2 \approx  4.558$
for $m_b=0$, and $\kappa=\pi^2\beta_b^\Gamma/16 
\approx 4.935$ for $m_f= 0$. 
(Note that
the result \eqref{proporo} appears to disagree with the estimates
$\zeta \sim \eta \left( v_s^2 -1/3\right)^2$
\cite{Hosoya:1983id,Horsley:1985dz}, later criticized in
\cite{Jeon:1995zm}.)

The paper is organized as follows.
In Section \ref{sec_geometry} we review the non-extremal
(finite-temperature) generalization of the Pilch-Warner flow. We
discuss holographic renormalization and thermodynamics of
$\caln=2^*$ gauge theory, and determine its equation of state to
leading order in $m_b/T\ll 1$, $m_f/T\ll 1$ in Section
\ref{sec_eq_state}. The equation of state determines the speed of
sound \eqref{csa}.
Computation of the sound attenuation constant requires evaluation
of the two-point correlation function of the stress-energy tensor,
which is a subject of Section \ref{sec_attenuation}. The
computation follows the general strategy of \cite{set}, albeit
with some technical novelties. We keep the parameters $m_f/T$ and
$m_b/T$ arbitrary until the very end, when we substitute into
equations the analytic solution for the metric valid in the
high-temperature regime only. Alas, even in that regime we have to
resort to numerical methods when computing the bulk viscosity.
Having found the sound wave pole of the stress-energy tensor
correlator, we confirm the result \eqref{csa} for the speed of
sound, and obtain the ratio of shear to bulk viscosity
\eqref{csax} for the strongly coupled $\caln=2^*$ gauge theory in
the high temperature regime. Our conclusions are presented in
Section \ref{sec_conclusions}. Details of the holographic
renormalization are discussed in Appendix \ref{holoren}. Some
technical details appear in Appendixes
\ref{appendix:coefficients_a}--\ref{appendix:structure_2}.

\section{Non-extremal $\caln=2^*$ geometry}
\label{sec_geometry}
The supergravity background dual to a finite
temperature $\caln=2^*$ gauge theory  \cite{bl} is a deformation
of the original $AdS_5\times S^5$ geometry induced by a pair of
scalars $\alpha$ and $\chi$ of the five-dimensional gauge
supergravity. (At zero temperature, such a deformation was
constructed by Pilch and Warner \cite{pw}.) According to the
general scenario of a holographic RG flow
\cite{Girardello:1998pd}, \cite{Distler:1998gb}, the asymptotic
boundary
 behavior of the supergravity scalars is related to the
bosonic and fermionic
mass parameters of the relevant operators inducing the RG flow
in the boundary gauge theory.
The action of the five-dimensional gauged supergravity is
\begin{equation}
\begin{split}
S=&\,
\int_{\calm_5} d\xi^5 \sqrt{-g}\ \call_5\\
=&\frac{1}{4\pi G_5}\,
\int_{\calm_5} d\xi^5 \sqrt{-g}\left[\ft14 R-3 (\del\a)^2-(\del\chi)^2-
\calp\right]\,,
\end{split}
\eqlabel{action5}
\end{equation}
where the potential\footnote{We set the five-dimensional gauged
supergravity coupling to one. This corresponds to setting the
radius $L$ of the five-dimensional sphere in the undeformed metric
to $2$.}
\begin{equation}
\calp=\frac{1}{16}\left[\frac 13 \left(\frac{\del W}{\del
\a}\right)^2+ \left(\frac{\del W}{\del \chi}\right)^2\right]-\frac
13 W^2\,
 \eqlabel{pp}
\end{equation}
is a function of  $\alpha$, $\chi$ determined by the
superpotential
\begin{equation}
W=- e^{-2\alpha} - \frac{1}{2} e^{4\alpha} \cosh(2\chi)\,.
\eqlabel{supp}
\end{equation}
The five-dimensional Newton's constant is
\begin{equation}
G_5\equiv \frac{G_{10}}{2^5\ {\rm vol}_{S^5}}=\frac{4\pi}{N^2}\,.
\eqlabel{g5}
\end{equation}
The action \eqref{action5} yields the Einstein equations
\begin{equation}
 R_{\mu\nu}=12 \del_\mu \a \del_\nu\a+ 4 \del_\mu\chi\del_\nu\chi
+\frac43 g_{\mu\nu} \calp\,,
\eqlabel{ee}
\end{equation}
as well as the equations for the scalars
\begin{equation}
\jsquare\alpha=\fft16\fft{\del\calp}{\del\alpha}\,,\qquad
\jsquare\chi=\fft12\fft{\del\calp}{\del\chi}\,.
\eqlabel{scalar}
\end{equation}
To construct a finite-temperature version of the Pilch-Warner
flow, one chooses an ansatz for the metric respecting rotational
but not the Lorentzian invariance\footnote{The full
ten-dimensional
 metric is given by Eq.~(4.12) in \cite{bl}.}
\begin{equation}
ds_5^2=-c_1^2(r)\ dt^2 +c_2^2(r)\ \left( d x_1^2 +  d x_2^2 + d x_3^2 \right)
+  dr^2\,.\\
\eqlabel{ab}
\end{equation}
The equations of motion for the background become
\begin{equation}
\begin{split}
&\a''+\a'\ \left(\ln{c_1c_2^3}\right)'
-\frac 16 \frac{\del\calp}{\del\a}=0\,,\\
&\c''+\c'\ \left(\ln{c_1c_2^3}\right)'-\frac 12 \frac{\del\calp}{\del\c}=0\,,\\
&c_1''+c_1'\ \left(\ln{c_2^3}\right)'+\frac 43c_1  \calp=0\,,\\
&c_2''+c_2'\ \left(\ln{c_1c_2^2}\right)'+\frac 43c_2  \calp=0\,,
\end{split}
\eqlabel{beom}
\end{equation}
where the prime denotes the derivative with respect to the radial
coordinate $r$.
 In addition, there is a first-order constraint
\begin{equation}
\left(\a'\right)^2+\frac 13 \left(\c'\right)^2 -\frac 13
\calp-\frac 12 (\ln c_2)'(\ln c_1 c_2)' =0\,. \eqlabel{backconst}
\end{equation}
A convenient choice of the radial coordinate is
\begin{equation}
x(r) = \frac{c_1}{c_2}\,, \;\;\; \;\;\;\;  x\in [0,1]\,.
\eqlabel{radgauge}
\end{equation}
With the new coordinate, the black brane's horizon is at $x=0$
while the boundary of the asymptotically $AdS_5$ space-time is at
$x=1$. The background equations of motion \eqref{beom} become
\begin{equation}
\begin{split}
&c_2''+4 c_2\ (\alpha')^2-\frac 1x c_2'-\frac{5}{c_2} (c_2')^2
+\frac 43 c_2\ (\c')^2 =0\,,\\
&\alpha''+\frac 1x \, \alpha'-
\frac{1}{12\, \calp c_2^2 x}\Biggl[       6 (\alpha')^2 c_2^2 x+
2 (\c')^2 c_2^2 x-3 c_2' c_2-6 (c_2')^2 x \Biggr] \,
 \frac{\del\calp}{\del\a} =0\,,\\
&\c''+\frac 1x\, \c'- \frac{1}{4\, \calp c_2^2  x}
\Biggl[ 6 (\alpha')^2 c_2^2 x+2 (\c')^2  c_2^2  x
-3 c_2' c_2-6 (c_2')^2 x \Biggr]\, \frac{\del\calp}{\del\c}=0\,,
\end{split}
\eqlabel{beomx}
\end{equation}
where the prime now denotes the derivative with respect to $x$.
A physical RG flow should correspond
 to the  background geometry with a regular horizon.
To ensure regularity, it is
 necessary to impose
the following boundary conditions at the horizon,
\begin{equation}
x\to 0_+ : \;\;\;\;\;\;\;
 \biggl\{\a(x),\ \c(x),\ c_2(x)\biggr\}\longrightarrow \biggl\{\dd_1,\dd_2,\dd_3\biggr\}\,,
\eqlabel{boundh}
\end{equation}
where $\dd_i$ are constants.
In addition,
the condition $\dd_3> 0$ guarantees the absence of
a naked singularity in the bulk.

The boundary conditions at $x= 1$ are determined from the
requirement that the solution should approach the $AdS_5$ geometry
as $x\to 1_-$:
\begin{equation}
x\to 1_- :  \;\;\;\;\;\;\;
\biggl\{\a(x),\ \c(x),\ c_2(x)\biggr\}\longrightarrow
\biggl\{0,0,\propto (1-x^2)^{-1/4}\biggr\}\,.
\eqlabel{bboun}
\end{equation}
The three supergravity parameters
$\dd_i$ uniquely determine a non-singular RG flow in the dual gauge theory.
 As we review shortly, they
are unambiguously related to the three physical parameters in
 the gauge theory:
the temperature $T$, and the bosonic and fermionic masses
 $m_b$, $m_f$ of the $\caln=2^*$ hypermultiplet components.

General analytical solution of the system
\eqref{beomx} with the boundary conditions
 \eqref{boundh}, \eqref{bboun} is unknown\footnote{One can study the system numerically, 
see Fig.~2 of \cite{bl}.}.
However, it is possible to find an analytical solution in the regime of
high temperatures.

\subsection{The high temperature Pilch-Warner flow}
\label{sec:HTPW}

Differential equations \eqref{beomx} describing finite temperature PW
renormalization group flow
admit a perturbative analytical solution
 at high temperature  \cite{bl}. The appropriate expansion parameters are
\begin{equation}
\dd_1\propto \left(\frac{m_b}{T}\right)^2\ll1,
\qquad \dd_2\propto \frac{m_f}{T}\ll 1\,.
\eqlabel{larget}
\end{equation}
Introducing a function $A(x)$ by
\begin{equation}
c_2\equiv e^A\,,
 \eqlabel{defc2}
\end{equation}
to leading nontrivial order in $\dd_1$, $\dd_2$  we have \cite{bl}
\begin{equation}
\begin{split}
A(x)=&\ln\dd_3-\frac 14\ \ln (1-x^2)+\dd_1^2\ A_1(x)+\dd_2^2\ A_2(x)\,,\\
\a(x)=&\dd_1\ \a_1(x)\,,\\
\c(x)=&\dd_2\ \c_2(x) \,,
\end{split}
\eqlabel{hightsol}
\end{equation}
where
\begin{equation}
\a_1=(1-x^2)^{1/2}\ _2 F_1\left(\ft 12,\ft 12; 1; x^2\right)\,,
\eqlabel{orko1}
\end{equation}
\begin{equation}
\c_2=(1-x^2)^{3/4}\ _2F_1\left(\ft 34, \ft 34; 1; x^2\right)\,,
\eqlabel{orko2}
\end{equation}
\begin{equation}
\begin{split}
A_1=&-4\int_0^x\ \frac{zdz}{(1-z^2)^2}
\left(\ga_1+\int_0^zdy\left(\frac{\del\a_1}{\del y}
\right)^2\frac{(1-y^2)^2}{y}\right)\,,\\
A_2=&-\frac 43\int_0^x\ \frac{zdz}{(1-z^2)^2}
\left(\ga_2+\int_0^zdy\left(\frac{\del\c_2}
{\del y}\right)^2\frac{(1-y^2)^2}{y}\right)\,.\\
\end{split}
\eqlabel{expsol}
\end{equation}
The constants $\ga_i$ were fine-tuned to satisfy the  boundary
conditions  \cite{bl}:
\begin{equation}
\ga_1=\frac{8-\pi^2}{2\pi^2},\qquad \ga_2=\frac{8-3\pi}{8\pi}\,.
\eqlabel{gai}
\end{equation}
The parameters $\dd_i$ are related to the parameters
$m_b$, $m_f$, $T$ of the dual gauge theory via
\begin{equation}
\begin{split}
\dd_1=&-\frac{1}{24\pi}\ \left(\frac{m_b}{T}\right)^2\,,\\
\dd_2=&\frac{\left[\Gamma\left(\ft 34\right)\right]^2}{2\pi^{3/2}}\ \frac{m_f}{T}\,,\\
2\pi T=&\dd_3\left(1+\frac{16}{\pi^2}\ \dd_1^2+\frac{4}{3\pi}\ \dd_2^2\right)\,.
\end{split}
\eqlabel{ddphys}
\end{equation}
Given the solution \eqref{hightsol}, $c_2$ is found from
Eq.~\eqref{defc2}, and $c_1 = x c_2$. The transition to the
original radial variable $r$ can be made by using the constraint
equation \eqref{backconst}. At ultra high temperatures
$\dd_1\rightarrow 0$,
 $\dd_2\rightarrow 0$, the conformal symmetry in the gauge theory
is restored, and one recovers the usual near-extremal black three-brane metric
\begin{equation}
ds_5^2 = (2\pi T)^2 (1-x^2)^{-1/2} \left( - x^2 dt^2 + d x_1^2 +
d x_2^2 + d x_3^2 \right) + \frac{dx^2}{(1-x^2)^2}\,,
\end{equation}
describing a gravity dual to a finite-temperature ${\cal N}=4$ SYM
in flat space.

\section{$\caln=2^*$ SYM equation of state and the speed of sound}
\label{sec_eq_state}

To determine the equation of state of $\caln=2^*$ gauge theory,
one needs to compute energy and pressure, given by the
corresponding one-point functions of the stress-energy tensor. In
computing the one-point functions, one has to deal with
divergences at the boundary of the asymptotically AdS space which
are related to UV divergences in the gauge theory. The method
addressing those issues is known as the holographic
renormalization \cite{hr1,hr2,hr3,hr4}.  Some
details of the holographic renormalization for the  $\caln=2^*$
gauge theory are given in Appendix \ref{holoren}. The method works
for arbitrary values of $m_b/T$, $m_f/T$, once the solution to
Eqs.~\eqref{beom} is known. In the high-temperature limit, energy
density and pressure can be computed explicitly
\begin{equation}
\begin{split}
\ee=&\frac 38 \pi^2 N^2 T^4
\left[ 1+\frac{64}{\pi^2}\
\left(\ln (\pi T)-1\right)\
 \delta_1^2-\frac {8}{3\pi}\ \delta_2^2\right]\,,\\
P=&\frac 18 \pi^2 N^2 T^4\left[ 1-\frac{192}{\pi^2}\ \ln(\pi T)\
\delta_1^2-\frac{8}{\pi}\ \delta_2^2\right]\,,
\end{split}
\eqlabel{finep}
\end{equation}
where $\delta_1$, $\delta_2$ are given by Eqs.~\eqref{ddphys}. One can
independently compute
 the entropy density of the non-extremal Pilch-Warner geometry \cite{bl}\,,
\begin{equation}
s=\frac 12 \pi^2 N^2 T^3\left(1-\frac{48}{\pi^2}\ \delta_1^2-\frac{4}{\pi}\ \delta_2^2\right)\,,
\end{equation}
and verify that the thermodynamic relation,
\begin{equation}
\cale-T s=-P\,,
\eqlabel{deff}
\end{equation}
is satisfied.
Alternatively, the free energy can be computed
 as a renormalized Euclidean action.
Then it can be shown  \cite{bh1}
  that the free energy density $\calf=-P$ obeys
$\calf=\cale-T s$ for arbitrary mass deformation
parameters $m_b/T$, $m_f/T$.
Finally, using Eq.~\eqref{ddphys} it can be verified that
\begin{equation}
d\cale=T ds\,.
\eqlabel{1stlaw}
\end{equation}
These checks demonstrate that the  $\caln=2^*$ thermodynamics is
unambiguously and correctly determined from gravity.

We can now evaluate leading correction to the speed of
sound in $\caln=2^*$ gauge theory plasma
at temperatures much larger than the conformal symmetry breaking
scales $m_f$, $m_b$.
Using Eqs.~\eqref{finep}, \eqref{ddphys} we find
\begin{equation}
v_s^2=\frac{\del P}{\del \ee}=\frac {\del P}{\del T}
\left(\frac{\del \ee}{\del T}\right)^{-1} =
\frac{1}{3}\left(1-\frac{64}{\pi^2}\ \dd_1^2-\frac{8}{3\pi}\
\dd_2^2\right)+\cdots\,,
\eqlabel{cs}
\end{equation}
where ellipses
 denote higher order terms in $m_f/T$ and $m_b/T$.
Substituting  $\delta_1$, $\delta_2$ from Eqs.~\eqref{ddphys}, we arrive at
 Eq.~\eqref{csa}\,.
In the next Section we confirm the result \eqref{cs} by evaluating
the two-point correlation function of the stress-energy tensor in
the sound mode channel and identifying the pole corresponding to
the sound wave propagation. In addition to confirming
Eq.~\eqref{cs}, this will allow us to compute the sound wave
attenuation constant and thus the bulk viscosity.

\section{Sound attenuation in $\caln=2^*$ plasma}
\label{sec_attenuation}

\subsection{Correlation functions from supergravity}

We calculate the poles of the two-point function of the stress-energy  tensor of the
${\caln}=2^*$
theory from gravity following the general scheme outlined in \cite{set}. 
Up to a certain index structure, the
generic thermal two-point function of stress-energy tensor is
determined by five scalar functions. In the hydrodynamic
approximation, one of these functions contains a pole at $\omega =
\omega(q)$ given by the dispersion relation \eqref{dispertion} and
corresponding to the sound wave propagation in $\caln=2^*$ plasma.
On the gravity side, the five functions characterizing the
correlator correspond to five gauge-invariant combinations of the
fluctuations of the gravitational background. The functions are
determined by the
 ratios of the connection coefficients of ODEs
satisfied by the gauge-invariant variables. Moreover, if one is
interested in poles rather than the full correlators, it is
sufficient to compute the quasinormal spectrum of the
corresponding gauge-invariant fluctuation. This approach is
illustrated in \cite{set} by taking ${\cal N}=4$ SYM as an
example. For a conformal theory such as ${\cal N}=4$ SYM, the
number of independent functions determining the correlator (and
thus the number of independent gauge-invariant variables on the
gravity side of the duality) is three. In the $\caln=2^*$ case,
the situation is technically more complicated, since we need to
take into account fluctuations of the two background scalars.
These matter fluctuations do not affect the "scalar" and the
"shear" channels\footnote{See \cite{set} for classification of
fluctuations.}, entering only the sound channel. In the sound
channel, this will lead to a system of three coupled ODEs for
three gauge-invariant variables mixing gravitational and scalar
fluctuations. The lowest or "hydrodynamic" quasinormal
frequency\footnote{Gapless frequency with the property
$\lim\limits_{q\rightarrow 0} \omega (q) = 0$ required by
hydrodynamics.} in the spectrum of the mode corresponding to the
sound wave gives the dispersion relation \eqref{dispertion} from
which the attenuation constant and thus the bulk viscosity can be
read off.

In this Section, we derive the equations for the gauge-invariant
variables for a generic finite-temperature Pilch-Warner flow, i.e.
without making any simplifying assumptions about the parameters of
the flow. Then we solve those equations in the regime $m_b/T\ll
1$, $m_f/T \ll 1$, where the background is known explicitly.
Solving the equations involves numerical integration. The final
result is given by Eqs.~\eqref{csa}, \eqref{csax}.

\subsection{Fluctuations of the non-extremal Pilch-Warner geometry}

Consider fluctuations of the background geometry
\begin{equation}
\begin{split}
g_{\mu\nu}&\to g^{BG}_{\mu\nu}+h_{\mu\nu}\,,\\
\a&\to \a_{BG}+\phi\,,\\
\c&\to \c_{BG}+\psi\,,
\end{split}
\eqlabel{fluctuations}
\end{equation}
where $g^{BG}_{\mu\nu}$
(more precisely, $c_1^{BG}$, $c_2^{BG}$), $\a_{BG}$,
$\c_{BG}$ are the solutions of the equations of motion
\eqref{beom}, \eqref{backconst}.
To simplify notations, in the following we use
$c_1$, $c_2$ to denote the background values of these fields,
omitting the label "BG".

For convenience, we partially fix the gauge by requiring
\begin{equation}
h_{tr}=h_{x_ir}=h_{rr}=0\,. \eqlabel{gaugec}
\end{equation}
This gauge-fixing is not essential, since we switch to
gauge-invariant variables shortly, but it makes the equations at
the intermediate stage less cumbersome. We orient the coordinate
system in such a way that the $x_3$ axis is directed along the
spatial momentum, and assume that
 all the fluctuations depend only on
$t,x_3,r$. The dependence of all variables on time and on the
spatial coordinate is of the form $\propto e^{-i\w t+iq x_3}$, so
the only non-trivial dependence is on the radial coordinate $r$.

The fluctuations can be classified according to their
transformation properties with respect to the $O(2)$ rotational
symmetry in the $x_1-x_2$ plane \cite{ne2}, \cite{set}. The set of
fluctuations corresponding to the sound wave mode consists of
\begin{equation}
h_{tt},\ h_{aa}\equiv h_{x_1x_1}+h_{x_2x_2},\ h_{tx_3},\
h_{x_3x_3},\ \phi ,\ \psi\,.
\end{equation}
Due to the $O(2)$ symmetry all other components of $h_{\mu\nu}$
can be consistently set to zero to linear order.
It will be convenient to introduce new variables $H_{tt}, H_{tz},
H_{aa}, H_{zz}$ by rescaling
\begin{equation}
h_{tt}=c_1^2\  H_{tt}\,,\;\;\;
h_{tz}=c_2^2\  H_{tz}\,, \;\;\;
h_{aa}=c_2^2\ H_{aa}\,,\;\;\;
h_{zz}=c_2^2\  H_{zz}\,.
\eqlabel{rescale}
\end{equation}
We also use $H_{ii}\equiv H_{aa}+H_{zz}$. Expanding
Eqs.~\eqref{beom}, \eqref{backconst} to linear order in
fluctuations, we obtain the coupled system of second-order ODEs
\begin{equation}
\begin{split}
H_{tt}'' &+H_{tt}'\ \left(\ln c_1^2c_2^3\right)'-H_{ii}'\ (\ln
c_1)' -\frac{1}{c_1^2}\left(q^2\frac{c_1^2}{c_2^2}\ H_{tt}+
\w^2\ H_{ii}+2\w q\ H_{tz}\right)\\
&-\frac 83  \left(\frac{\del\calp}{\del\a}\ \aa1+\frac
{\del\calp}{\del\c}\ \cc1\right) = 0\,,
\end{split}
\eqlabel{fl1}
\end{equation}
\begin{equation}
\begin{split}
H_{tz}'' &+H_{tz}'\ \left(\ln\frac{c_2^5}{c_1}\right)'
+\frac{1}{c_2^2}\ \w q\ H_{aa} =0\,,
\end{split}
\eqlabel{fl2}
\end{equation}
\begin{equation}
\begin{split}
H_{aa}'' &+ H_{aa}'\ \left(\ln c_1c_2^5\right)'+(H_{zz}'-H_{tt}')\
(\ln c_2^2)'
+\frac{1}{c_1^2}\left(\w^2-q^2\frac{c_1^2}{c_2^2}\right)\ H_{aa}\\
&+\frac {16}{3} \left(\frac{\del\calp}{\del\a}\
\aa1+\frac{\del\calp}{\del\c}\ \cc1\right)=0\,,
\end{split}
\eqlabel{fl3}
\end{equation}
\begin{equation}
\begin{split}
H_{zz}'' &+H_{zz}'\ \left(\ln c_1c_2^4\right)'+\left(H_{aa}'-H_{tt}'\right)\
(\ln c_2)'
+\frac{1}{c_1^2}\Biggl[ \w^2\ H_{zz}+2\w q\ H_{tz} \\
&+q^2\frac{c_1^2}{c_2^2}\left( H_{tt}-H_{aa} \right)\Biggr]+\frac {8}{3}
\left(\frac{\del\calp}{\del\a}\ \aa1+\frac{\del\calp}{\del\c}\
\cc1\right)=0\,,
\end{split}
\eqlabel{fl4}
\end{equation}
\begin{equation}
\begin{split}
\aa1''&+\aa1'\ \left(\ln c_1c_2^3\right)'+\frac 12 \a_{BG}'\
\left(H_{ii}-H_{tt}\right)'+\frac{1}{c_1^2}
\left(\w^2-q^2\frac{c_1^2}{c_2^2}\right)\ \aa1\\
&-\frac {1}{6} \left(\frac{\del^2\calp}{\del\a^2}\
 \aa1+\frac{\del^2\calp}{\del\a\del\c}\ \cc1\right)=0\,,
\end{split}
\eqlabel{fl5}
\end{equation}
\begin{equation}
\begin{split}
\cc1'' &+\cc1'\ \left(\ln c_1c_2^3\right)'+\frac 12 \c_{BG}'\
\left( H_{ii}-H_{tt}\right)'+\frac{1}{c_1^2}
\left(\w^2-q^2\frac{c_1^2}{c_2^2}\right)\ \cc1\\
&-\frac {1}{2}  \left(\frac{\del^2\calp}{\del\a\del\c}\
\aa1+\frac{\del^2\calp}{\del\c^2}\ \cc1\right) =0\,,
\end{split}
\eqlabel{fl6}
\end{equation}
where the prime denotes the derivative with respect to $r$, and
all the derivatives of the potential $\calp$ are evaluated in the
background geometry. In addition, there are three first-order
equations
\begin{equation}
\begin{split}
\,& \w\left[ H_{ii}'+\left(\ln\frac{c_2}{c_1}\right)'\
H_{ii}\right]+q\left[ H_{tz}'+2\left(\ln\frac{c_2}{c_1}\right)'\
H_{tz}\right] \\ &+8\w\ \left( 3 \a_{BG}'\ \aa1+ \c_{BG}'\
\cc1\right) =0\,,
\end{split}
\eqlabel{const1}
\end{equation}
\begin{equation}
\begin{split}
&q\left[ H_{tt}'-\left(\ln\frac{c_2}{c_1}\right)'\
H_{tt}\right]+\frac{c_2^2}{c_1^2}\w\ H_{tz}'-q\ H_{aa} -8q\
\left(3 \a_{BG}'\ \aa1+\c_{BG}'\ \cc1\right)=0\,,
\end{split}
\eqlabel{const2}
\end{equation}
\begin{equation}
\begin{split}
&(\ln c_1c_2^2)' \, H_{ii}'-(\ln{c_2^3})'\ H_{tt}'+\frac{1}{c_1^2}
\left[\w^2\ H_{ii}+2\w q\ H_{tz}+q^2\
\frac{c_1^2}{c_2^2}\left(H_{tt}-H_{aa}\right)\right]\\
&+4 \left(\frac{\del\calp}{\del\a}\ \aa1+\frac{\del\calp}{\del\c}\
\cc1\right)-8\ \left( 3\a_{BG}'\ \aa1'+\c_{BG}'\
\cc1'\right)=0\,.
\end{split}
\eqlabel{const3}
\end{equation}

\subsection{Gauge-invariant variables}

A convenient way to deal with the fluctuation equations is to
introduce gauge-invariant variables \cite{set}. (Such an approach
has long been used in cosmology \cite{bar}, \cite{bar1} and in
studying black hole fluctuations \cite{ik}.) Under the
infinitesimal diffeomorphisms
\begin{equation}
x^\mu \to x^\mu+\xi^\mu\\
\end{equation}
the metric and the scalar field fluctuations transform as
\begin{equation}
\begin{split}
g_{\mu\nu}&\to g_{\mu\nu}-\nm\xi_\nu-\nn\xi_\mu\,,\\
\phi &\to \phi - \nabla^\lambda\alpha_{BG}\ \xi_\lambda\,, \\
\psi &\to \psi - \nabla^\lambda\chi_{BG}\ \xi_\lambda\,,
\end{split}
\eqlabel{diffeo}
\end{equation}
where the covariant derivatives are computed in the background metric.
One finds the following linear combinations of fluctuations
 which are invariant under the diffeomorphisms \eqref{diffeo}
\begin{equation}
\begin{split}
Z_H=&4\frac{q}{\w} \ H_{tz}+2\
H_{zz}-H_{aa}\left(1-\frac{q^2}{\w^2}\frac{c_1'c_1}{c_2'c_2}\right)+2\frac{q^2}{\w^2}
\frac{c_1^2}{c_2^2}\ H_{tt}\,,\\
Z_\phi = &\phi-\frac{\alpha_{BG}'}{(\ln c_2^4)'}\ H_{aa}\,,\\
Z_\psi = &\psi-\frac{\chi_{BG}'}{(\ln c_2^4)'}\ H_{aa}\,.
\end{split}
\eqlabel{physical}
\end{equation}
Using Eqs.~\eqref{fl1}-\eqref{const3},
one finds the new variables $Z_H$, $Z_{\phi}$, $Z_{\psi}$
satisfy the following system of coupled equations
\begin{subequations}\label{za}
\begin{eqnarray}
\,&&A_H \,  Z_H'' + B_H \, Z_H' + C_H \,  Z_H + D_H \,
 Z_{\phi} + E_H \,  Z_{\psi} =0\,,  \\
\,&&A_\phi   Z_\phi '' + B_\phi  Z_\phi' + C_\phi   Z_\phi
 + D_\phi  Z_{\psi} + E_\phi   Z_{H}' + F_\phi  Z_H =0\,,  \\
\,&&A_\psi   Z_\psi '' + B_\psi  Z_\psi' + C_\psi   Z_\psi
 + D_\psi  Z_{\phi} + E_\psi   Z_{H}' + F_\psi  Z_H =0\,,
\end{eqnarray}
\end{subequations}
where the coefficients depend on the background values $c_1$,
$c_2$, $\alpha_{BG}$, $\chi_{BG}$. (The coefficients are given
explicitly in Appendixes \ref{appendix:coefficients_a},
\ref{appendix:coefficients_b}, \ref{appendix:coefficients_c}.)
Eqs.~\eqref{za} describe fluctuations of the background for
arbitrary values of the deformation parameters $m_b/T$, $m_f/T$.

The analysis of Eqs.~\eqref{za} is simplified by switching to the
new radial coordinate  \eqref{radgauge}. Asymptotic behavior of
the solutions to  Eqs.~\eqref{za} near the horizon, $x\to 0_+$,
corresponds to waves incoming to the horizon and outgoing from it,
i.e. for each of the gauge-invariant variables we have $Z_H,
Z_\phi, Z_\psi \propto x^{\pm i \w}$. We are interested in the
lowest quasinormal frequency of the ``sound wave'' variable $Z_H$.
To ensure that this frequency is indeed the hydrodynamic
dispersion relation \eqref{dispertion}
 appearing as
the pole in the retarded two-point function of the stress-energy tensor
of the ${\cal N}=2^*$ SYM, one has to impose the following boundary
conditions \cite{set}
\nxt the incoming wave boundary condition on all fields at the horizon:
$Z_H, Z_\phi, Z_\psi \propto x^{- i \w}$ as $x\to 0_+$;
\nxt  Dirichlet condition on $Z_H$ at the boundary $x=0$: $Z_H(0)=0$.

The incoming boundary condition on physical modes implies that
\begin{equation}
\begin{split}
Z_H(x)=x^{-i\w} \tilde{Z}_H (x),\qquad Z_\phi(x)=x^{-i\w} \tilde{Z}_\phi (x),
\qquad Z_\psi(x)=x^{-i\w} \tilde{Z}_\psi (x)\,,
\label{axxa}
\end{split}
\end{equation}
where  $\tilde{Z}_H$,  $\tilde{Z}_\phi$, $\tilde{Z}_\psi$ are
 are regular functions at the horizon.
 Without the loss of generality the integration constant can be fixed as
\begin{equation}
\tilde{Z}_H \bigg|_{x\to 0_+}=1\,.
\eqlabel{bconditions}
\end{equation}
Then the dispersion relation \eqref{dispertion}
is  determined by the Dirichlet condition at the boundary \cite{set}
\begin{equation}
\tilde{Z}_H (x)\bigg|_{x\to 1_-}=0\,.
\eqlabel{poledisp}
\end{equation}
Following \cite{Policastro:2001yb}, \cite{ne2},
a solution to Eqs.~\eqref{za} can in principle be found
in the hydrodynamic approximation
 as a series in small $\w$, $q$ (more precisely, $\w/T \ll 1$,
$q/T\ll 1$), provided the background values $c_1$, $c_2$,
$\alpha_{BG}$, $\chi_{BG}$ are known explicitly.
Here we consider the high-temperature limit \eqref{larget} discussed in
Section \ref{sec:HTPW},
and expand all fields in series for $\dd_1 \ll 1$, $\dd_2 \ll 1$.
Accordingly, we introduce
\begin{equation}
\begin{split}
&\tilde{Z}_H =\biggl(Z^{0}_0+
\dd_1^2\ Z^{0}_1+\dd_2^2\ Z^{0}_2\biggr)+ i q\
 \biggl(Z^{1}_0+\dd_1^2\ Z^{1}_1+\dd_2^2\ Z^{1}_2\biggr)\,,\\
&\tilde{Z}_\phi =\dd_1 \biggl(Z_\phi^0+i q\ Z_\phi^1\biggr)\,,\\
& \tilde{Z}_\psi   =\dd_2 \biggl(Z_\psi^0+ i q\ Z_\psi^1\biggr)\,,
\end{split}
\eqlabel{defzz}
\end{equation}
where the upper index refers to either the leading, $\propto q^0$,
or to the next-to-leading, $\propto q^1$ order in the hydrodynamic
approximation, and the lower index keeps track of the bosonic,
$\dd_1$, or fermionic, $\dd_2$, mass deformation parameter.
Eqs.~\eqref{axxa}, \eqref{defzz} represent a perturbative solution
of
 Eqs.~\eqref{za} to first order in $\w$, $q$, and to leading nontrivial order
in $\dd_1$, $\dd_2$.
From \eqref{bconditions}, the boundary conditions at the
horizon are
\begin{equation}
\begin{split}
&Z^0_0\bigg|_{x\to 0_+}=1,\qquad Z^0_1\bigg|_{x\to 0_+}=0,
\qquad Z^0_2\bigg|_{x\to 0_+}=0\,,\\
&Z^1_0\bigg|_{x\to 0_+}=0,\qquad Z^1_1\bigg|_{x\to 0_+}=0,
\qquad Z^1_2\bigg|_{x\to 0_+}=0\,.\\
\end{split}
\eqlabel{hb1}
\end{equation}
The Dirichlet condition at the boundary \eqref{poledisp} becomes
\begin{equation}
\begin{split}
&Z^0_0\bigg|_{x\to 1_-}=0, \qquad Z^0_1\bigg|_{x\to 1_-}=0,
\qquad Z^0_2\bigg|_{x\to 1_-}=0\,,\\
&Z^1_0\bigg|_{x\to 1_-}=0,\qquad Z^1_1\bigg|_{x\to 1_-}=0,
\qquad Z^1_2\bigg|_{x\to 1_-}=0\,.\\
\end{split}
\eqlabel{poledisp1}
\end{equation}
We also find it convenient to parametrize the frequency as
\begin{equation}
\w=\frac{q}{\sqrt{3}}\biggl(1+\b_1^v\ \dd_1^2+\b_2^v\ \dd_2^2\biggr)- \frac {i q^2}{3}
\biggl(1+\b_1^\G\ \dd_1^2+\b_2^\G\ \dd_2^2\biggr)\,.
\eqlabel{disprel}
\end{equation}
In the absence of mass deformation ($\dd_1=\dd_2=0$),
Eq.~\eqref{disprel} reduces to the sound wave dispersion relation
 for the
 $\caln=4$ SYM plasma  \cite{ne4}. The parametrization \eqref{disprel}
reflects our expectations that the  conformal $\caln=4$ SYM dispersion relation
will be modified by corrections proportional to the mass deformation parameters
$\dd_1$, $\dd_2$. Our goal is to determine the coefficients
$\b_1^v$, $\b_2^v$, $\b_1^\G$,  $\b_2^\G$ by requiring that the
perturbative solution  \eqref{defzz} should satisfy the boundary conditions
 \eqref{hb1}, \eqref{poledisp1}.

Using the high-temperature non-extremal
 Pilch-Warner  background \eqref{hightsol},
parameterizations
\eqref{defzz} and \eqref{disprel},
and rewriting  Eqs.~\eqref{za} in the radial coordinate  \eqref{radgauge}, we
obtain three sets of ODEs describing, correspondingly:
\nxt the pure $\caln=4$ physical sound wave mode ($\dd_1=0$, $\dd_2=0$);
\nxt corrections to pure $\caln=4$ physical sound wave mode due
 to the bosonic mass deformation ($\dd_1\neq 0$, $\dd_2=0$) ;
\nxt corrections to pure $\caln=4$ physical sound wave mode
 due to fermionic mass deformation ($\dd_1 = 0$, $\dd_2 \neq 0$).

In the remaining part of this subsection we derive equations
corresponding to each of these three sets.

\subsubsection{Sound wave quasinormal mode for $\caln=4$ SYM}
Setting $\dd_1=\dd_2=0$ in  Eqs.~\eqref{za},
\eqref{defzz}, \eqref{disprel} leads to the following
equations
\begin{equation}
x(x^2+1) \frac{d^2 Z^0_0}{d x^2} + (1-3x^2)\ \frac{d Z_0^0}{d x}
+4x Z_0^0 =0\,,
\eqlabel{n4sw0}
\end{equation}
\begin{equation}
\begin{split}
&x (x^2+1)^2\ \frac{ d^2 Z^1_0}{d x^2}
- (x^2+1) (3 x^2-1)\  \frac{d Z^1_0}{d x} +4  x (x^2+1)\ Z^1_0\\
&-\frac{2}{\sqrt{3}}
(x^2-1)^2\ \frac{ d Z^0_0}{d x} +\frac{4}{\sqrt{3}} x(x^2-1)\ Z^0_0 =0\,.
\end{split}
\eqlabel{n4sw1}
\end{equation}
The general solution of Eq.~\eqref{n4sw0} is
\begin{equation}
Z^0_0=\calc_1 (1-x^2)+\calc_2 \left[ (x^2-1)\ln x -2\right]\,,
\eqlabel{solz00}
\end{equation}
where $\calc_1$, $\calc_2$ are integration constants. The
condition of
 regularity at the horizon and the boundary condition \eqref{hb1}
lead to
\begin{equation}
Z^0_0=1-x^2\,.
\eqlabel{z00fin}
\end{equation}
Notice that the boundary condition \eqref{poledisp1} is
automatically satisfied, as it should be, since \eqref{disprel}
with $\dd_1=0$, $\dd_2=0$ is the correct quasinormal frequency of
the gravitational fluctuation in the sound wave channel for the
background dual to pure  $\caln=4$ SYM plasma. Given the solution
\eqref{z00fin}, the general solution to Eq.~\eqref{n4sw1} reads
\begin{equation}
Z^1_0=\calc_3 (1-x^2)+\calc_4 \left[ (x^2-1)\ln x -2\right]\,,
\eqlabel{solz10}
\end{equation}
where $\calc_3$, $\calc_4$ are integration constants. Imposing
regularity at the horizon and the boundary condition \eqref{hb1}
gives
\begin{equation}
Z^1_0=0\,.
\eqlabel{z10fin}
\end{equation}
Again, the boundary condition \eqref{poledisp1} at $x=0$
is automatically satisfied as a result of the parametrization \eqref{disprel}.

\subsubsection{Bosonic mass deformation of the  $\caln=4$ sound wave mode}
Turning on the bosonic mass deformation parameter $\dd_1$
(while keeping  $\dd_2=0$)
 and
using the zeroth-order solutions \eqref{z00fin}, \eqref{z10fin}, we find
from  Eqs.~\eqref{za}, \eqref{defzz}, \eqref{disprel}
\begin{equation}
\begin{split}
&3 x^3 (x^2-1)^2\ \frac{d^2 Z_\phi^0}{d x^2} + 3 x^2 (x^2-1)^2\
\frac{d Z_\phi^0}{d x} + 3 x^3\  Z_\phi^0+2 (x^2-1)^2\ \frac{ d \a_{1}}{d x}\\
&-x (x^2-1)\ \a_{1}=0\,,
\end{split}
\eqlabel{za0}
\end{equation}

\begin{equation}
\begin{split}
&x^2(x^4-1)\ \frac{d^2 Z_1^0}{d x^2} -x(x^2-1)(3x^2-1)\ \frac{ d Z_1^0}{ d x}
+4x^2(x^2-1)\ Z_1^0\\
&+192x^2\ \left[4x(1-x^2)\ \frac{d \alpha_{1}}{d x}
+(1+x^2)\ \alpha_{1} \right]\ Z_\phi^0+16x(x^2-1)^3\
\frac{d A_1}{d x}\\
&+32(x^4-1)(x^2-1)^2\ \left(\frac{d \alpha_{1}}{d x}\right)^2
+8x^2(x^2-1)\ \b_1^v =0\,,
\end{split}
\eqlabel{z01}
\end{equation}

\begin{equation}
\begin{split}
&3 x^3 (x^2-1)^2\ \frac{d^2 Z_\phi^1}{d x^2}
+3 x^2 (x^2-1)^2\ \frac{d Z_\phi^1}{d x} +3 x^3\ Z_\phi^1\\
&-2 \sqrt{3} x^2 (x^2-1)^2\ \frac{d Z_\phi^0}{d x}
-\sqrt{3} (x^2-1)\ \left[ 2  (x^2-1) \frac{d \alpha_1}{d x}
 -x\ \alpha_1\right] =0\,,
\end{split}
\eqlabel{za1}
\end{equation}

\begin{equation}
\begin{split}
&x^2 (x^4-1) (1+x^2)\ \frac{d^2 Z_1^1}{d x^2}
-x (x^4-1) (3 x^2-1)\ \frac{d Z_1^1}{d x} +4 x^2 (x^4-1)\ Z_1^1\\
&-192 x^2 (1+x^2)\ \left[ 4 x(x^2-1)\ \frac{d \alpha_1}{ d x}
-(x^2+1)\ \alpha_1\right]  Z_\phi^1\\
&-\frac{2}{\sqrt{3}} x (x^2-1)^3\ \frac{d Z_1^0}{d x}+\frac{4}{\sqrt{3}} x^2
 (x^2-1)^2\ Z_1^0\\
&-128 \sqrt{3} x^2\ \left[2x(x^2-1)(3x^2+1)\
\frac{d \alpha_1}{ d x}-(1+x^2)^2\ \alpha_1
\right]\ Z_\phi^0\\
&-\frac{32}{\sqrt{3}} (1+x^2)^2 (x^2-1)^3\
\left( \frac{ d \alpha_1}{d x}\right)^2-\frac{16}{\sqrt{3}}
 x (x^2+3) (x^2-1)^3\ \frac{d A_1}{ d x}\\
&-\frac{8}{\sqrt{3}} (x^2+3) (x^2-1) x^2\ \b_1^v-\frac{8}{\sqrt{3}}
  x^2(x^4-1)\ \b_1^\Gamma =0\,,
\end{split}
\eqlabel{z11}
\end{equation}
where  functions $A_1(x)$, $\alpha_1(x)$ are given by
Eqs.~\eqref{orko1}, \eqref{expsol}.

\subsubsection{Fermionic mass deformation of the $\caln=4$ sound wave mode}
Similarly, turning on the fermionic mass deformation parameter
$\dd_2$ and leaving  $\dd_1=0$ we get
\begin{equation}
\begin{split}
&12 x^3 (x^2-1)^2\ \frac{d^2 Z_\psi^0}{d x^2}
+12 x^2 (x^2-1)^2\ \frac{ d Z_\psi^0}{d x} +9 x^3\ Z_\psi^0+8 (x^2-1)^2\
 \frac{d \chi_2}{d x}\\
&-3 x (x^2-1)\  \chi_2 =0\,,
\end{split}
\eqlabel{zc0}
\end{equation}

\begin{equation}
\begin{split}
&3 x^2 (x^4-1)\  \frac{ d^2 Z_2^0}{d x^2}
-3 x (x^2-1) (3 x^2-1)\ \frac{ d Z_2^0}{ d x} +12 x^2 (x^2-1)\ Z_2^0\\
&-48 x^2 \left[ 16 x (x^2-1)\ \frac{d \chi_2}{ d x}
-3 (1+x^2)\ \chi_2 \right]\ Z_\psi^0\\
&+32 (x^4-1) (x^2-1)^2\ \left(\frac{ d \chi_2}{ d x }
\right)^2+48 x (x^2-1)^3\ \frac{ d A_2}{ d x } +24 x^2 (x^2-1)\ \b_2^v =0 \,,
\end{split}
\eqlabel{z20}
\end{equation}

\begin{equation}
\begin{split}
&12 x^3 (x^2-1)^2\  \frac{d^2 Z_\psi^1}{d x^2}
+24 x^4 (x^2-1)\ \frac{d Z_\psi^1}{d x}
+9 x^3\ Z_\psi^1\\
&-8 \sqrt{3} x^2 (x^2-1)^2\ \frac{d Z_\psi^0}{d x}
- \sqrt{3} (x^2-1)\left[ 8(x^2-1)\ \frac{d \chi_2}{d x} -3 x\
\chi_2\right] =0\,,
\end{split}
\eqlabel{zc1}
\end{equation}
\begin{equation}
\begin{split}
&3 x^2 (x^4-1) (1+x^2)\ \frac{d^2 Z_2^1}{ d x^2}
-3 x (x^4-1) (3 x^2-1)\ \frac{d Z_2^1}{d x}
+12 x^2 (x^4-1)\ Z_2^1\\
&-48 x^2 (1+x^2) \left[ 16 x (x^2-1) \frac{ d \chi_2}{d x}
-3 (1+x^2) \chi_2\right] Z_\psi^1\\
&-2 \sqrt{3} x (x^2-1)^3\ \frac{ d Z_2^0}{ d x}
+4 \sqrt{3} x^2 (x^2-1)^2\ Z_2^0\\
&-32 \sqrt{3} x^2 \left[ 8 x (x^2-1) (3 x^2+1)
\frac{d \chi_2}{ d x }-3 \chi_2 (1+x^2)^2\right]\ Z_\psi^0\\
&-\frac{32}{\sqrt{3}} (1+x^2)^2 (x^2-1)^3\
 \left(\frac{d \chi_2}{d x}\right)^2-16 \sqrt{3} x (x^2+3) (x^2-1)^3\
 \frac{ d A_2}{d x} \\
&-8 \sqrt{3} (x^2+3) (x^2-1) x^2\ \b_2^v-8 \sqrt{3} x^2(x^4-1)\ \b_2^\Gamma =0\,,
\end{split}
\eqlabel{z21}
\end{equation}
where  functions $A_2(x)$, $\chi_2 (x)$ are given by
Eqs.~\eqref{orko2}, \eqref{expsol}.

\section{Solving the fluctuation equations}

In this Section, we provide some details on solving the boundary
value problems for the bosonic and fermionic mass deformations of
the  $\caln=4$ SYM sound wave mode, discussing in particular the
numerical techniques involved. We start with solving
Eqs.~\eqref{zc0}--\eqref{z21} subject to the boundary conditions
\eqref{hb1}--\eqref{poledisp1}.
\subsection{Speed of sound and   attenuation constant
to $\calo\left(\dd_2^2\right)$ in $\caln=2^* $ plasma}
Here we solve Eqs.~\eqref{zc0}--\eqref{z21} subject to the boundary conditions
\eqref{hb1}--\eqref{poledisp1}.
Notice that the coefficient $\b_2^v$
can be determined by imposing
 the boundary condition \eqref{poledisp1} on the
perturbation mode $Z_2^0$. The coefficient
$\b_2^\G$ can then be extracted by solving for the  mode $Z_2^1$
subject to
 the  boundary condition \eqref{poledisp1}.

For the purposes of
numerical analysis  it will be convenient to redefine the radial coordinate
by introducing
\begin{equation}
y\equiv \frac{x^2}{1-x^2}\,.
\eqlabel{redef}
\end{equation}
Near the horizon ($x=0$)
we have $y=x^2+\calo(x^4)$, while the boundary ($x=1$)
is pushed to $y\to +\infty$.
The computation proceeds in four steps:
\nxt First, we solve Eq.~\eqref{zc0}.
Applying the arguments of
 \cite{set} to  Eq.~\eqref{zc0}, we find that the appropriate
boundary condition on $Z_\psi^0$ at $y\rightarrow \infty$ is
\begin{equation}
Z_\psi^0\sim y^{-3/4} \qquad {\rm as}\qquad  y\to \infty\,.
\eqlabel{zc0b}
\end{equation}
Eq.~\eqref{zc0b}, along with the requirement of regularity at the horizon,
 uniquely determines the solution $Z_\psi^0(x)$.
\nxt Second, we solve Eq.~\eqref{z20}. The solution is an analytic
 expression involving integrals of the solution  $Z_\psi^0(x)$ constructed in Step 1.
 Again, the regularity at the horizon plus the horizon boundary
condition \eqref{hb1}  uniquely determine $Z_2^0(x)$.
The coefficient $\b_2^v$ is evaluated numerically after imposing the
boundary condition \eqref{poledisp1}.
\nxt Third, we solve Eq.~\eqref{zc1}. The boundary condition
 \begin{equation}
Z_\psi^1\sim y^{-3/4} \qquad {\rm as}\qquad  y\to \infty\,
\eqlabel{zc1b}
\end{equation}
uniquely determines the solution.
\nxt Finally, we solve  Eq.~\eqref{z21}.
The regularity at the horizon and the horizon boundary
condition \eqref{hb1}  uniquely determine $Z_2^1(x)$.
Then the coefficient $\b_2^\G$ is determined numerically
by imposing the boundary condition \eqref{poledisp1}.

Having outlined the four-step approach, we now provide more details on
each of the steps involved.

\subsubsection{Step 1}
Asymptotic behavior near the boundary of the general solution to
Eq.~\eqref{zc0} regular at the horizon is given by
\begin{equation}
Z_\psi^0= {\cal A}_\psi^0 y^{-1/4} + \cdots + {\cal B}_\psi^0 y^{-3/4} + \cdots\,,
\eqlabel{ozc0}
\end{equation}
where  ${\cal A}_\psi^0$, ${\cal B}_\psi^0$ are the connection coefficients of the ODE.
Rescaling the dependent variable as
\begin{equation}
Z_\psi^0 \equiv (1+y)^{-3/4}\ g_\psi (y)\,,
\eqlabel{defgc}
\end{equation}
we find that the new function  $g_\psi (y)$
 satisfies the following differential equation
\begin{equation}
\begin{split}
&\frac{d^2 g_\psi}{d y^2}
+ \frac{y+2}{2y(1+y)}\ \frac{d g_\psi}{d y} -\frac{9\, g_\psi }{16y(1+y)^2}\
-\frac{3}{512y(1+y)^2}\ _2F_1\left(\frac 74, \frac 74; 3; \frac{y}{1+y}\right)=0\,.
\end{split}
\eqlabel{gceq}
\end{equation}
Imposing the regularity condition at the horizon, one constructs a
   power series solution near $y=0$
\begin{equation}
g_\psi =g^0_\psi+\left(\frac{9}{16} g^0_\psi+\frac{3}{512}\right)
 y+\left(-\frac{135}{1024} g^0_\psi +\frac{1}{8192}\right) y^2
+\calo(y^3)\,.
\eqlabel{gc1p}
\end{equation}
The integration constant $g^0_\psi$ is
fixed by requiring that, as Eq.~\eqref{zc0b} suggests,
\begin{equation}
g_\psi\sim \calo(1)\qquad {\rm as}\qquad y\to \infty\,.
\eqlabel{bbgc}
\end{equation}
For the numerical analysis, we had constructed
 the power series solution \eqref{gc1p} to order
$\calo(y^{19})$. Then, for a given $g^0_\psi$, we numerically
integrated $g_\psi$ from some small initial value  $y=y_{in}$,
using the constructed power series solution to set the initial
values of $g_\psi(y_{in})$, $g_\psi'(y_{in})$. (This is necessary
as the differential
 equation has a singularity at $y=0$.)
 We verified that
the final numerical result is insensitive to the choice of  $y_{in}$
as long as  $y_{in}$ is sufficiently small (we used  $y_{in}=10^{-15}$).
Then we applied a ``shooting'' method to determine $g^0_{\psi}$. The
 ``shooting'' method is convenient in view of the asymptotic
behavior \eqref{ozc0} near the boundary:
unless $g^{0}_\psi$
 is fine-tuned appropriately,
for large values of $y$, $g_\psi(y)$ would diverge, $g_\psi\propto y^{1/2}$
as $y\to \infty$.
We thus find
\begin{equation}
g_\psi^0 \approx -0.02083333(4)\,, \eqlabel{g0cres}
\end{equation}
where the brackets indicate an error in the corresponding digit.
We conjecture that the exact result is
\begin{equation}
g_\psi^0=-\frac{1}{48}\,.
\eqlabel{g0cfin}
\end{equation}
(Note that a formal analytical solution to 
Eq.~\eqref{gceq} is available. It allows to express
$g_\psi^0$ in terms of a certain  definite integral. Numerical
 evaluation of the integral confirms the result \eqref{g0cfin}.)

\subsubsection{Step 2}
The  solution to Eq.~\eqref{z20} regular at the horizon and
satisfying the boundary condition \eqref{hb1} is
\begin{equation}
Z_2^0(x)=\frac{8}{3} (x^2-1)\ \cali_{Z_2^0}^a(x)
-\frac{8}{3}\left[(x^2-1)\ln x-2\right] \ \cali_{Z_2^0}^b(x)\,,
\eqlabel{z20s}
\end{equation}
where
\begin{equation}
\begin{split}
\cali_{Z_2^0}^a(x)=&-\frac{3x^2(x^2+3+2\ln x)}{4(1+x^2)^2}\ \beta_2^v
+\int_0^x\ dz\ \frac{(z^2-1)\ln z-2}{z(z^4-1)(z^2+1)^2}\ \times \\
&\biggl\{
6z^2\left[-16z(z^2-1)\ \frac{d \chi_2}{d z}
+3(z^2+1)\ \chi_2\right]\ Z_\psi^0 (z)\\
&+4(z^4-1)(z^2-1)^2\ \left(\frac{d \chi_2}{d z}\right)^2
+6z(z^2-1)^3\ \frac{d A_2}{d z} \biggr\}\,,
\end{split}
\eqlabel{z20a}
\end{equation}
\begin{equation}
\begin{split}
\cali_{Z_2^0}^b(x)=&-\frac{3x^2}{2(1+x^2)^2}\ \beta_2^v
+\int_0^x\ dz\ \frac{1}{z(z^2+1)^3}\ \times \\
&\biggl\{
6z^2\left[
-16z(z^2-1)\ \frac{d\chi_2}{d z}+3(z^2+1)\ \chi_2\right]\ Z_\psi^0 (z)\\
&+4(z^2-1)(z^2-1)^2\ \left(\frac{d \chi_2}{d z}\right)^2+6z(z^2-1)^3\
 \frac{ d A_2}{d z}\,.
\biggr\}
\end{split}
\eqlabel{z20b}
\end{equation}
We explicitly verified that
\begin{equation}
\lim_{x\to 1_-} \cali_{Z_2^0}^a(x)= \calo(1)\,.
\eqlabel{checkz20a}
\end{equation}
Thus the boundary condition \eqref{poledisp1} becomes
\begin{equation}
\lim_{x\to 1_-} \cali_{Z_2^0}^b(x)= 0\,.
\eqlabel{checkz20b}
\end{equation}
Numerically solving Eq.~\eqref{checkz20b} for $\b_2^v$, we find
\begin{equation}
\b_2^v \approx -\frac {4}{3\pi}\ \times\ 0.9999(5)\,,
\eqlabel{b1vnumer}
\end{equation}
where we factored out the value $-4/3\pi$ for the coefficient
$\b_2^v$ obtained earlier from the
equation of state (see Eq.~\eqref{cs}).

\subsubsection{Step 3}
The general solution to Eq.~\eqref{zc1} has the form
\begin{equation}
\begin{split}
Z_\psi^1=&\sin\left(
\frac{\sqrt{3}}{2} \arctanh\  x\right)\times \biggl\{
\calc_1-\frac 16 \int_0^x dz
\frac{\cos\left( \frac{\sqrt{3}}{2} \arctanh\  z\right)}{z^3}\ \times\\
&\left(8z^2(z^2-1)\ \frac{d Z_\psi^0}{d z}
+8(z^2-1)\ \frac{d \c_2}{d z} -3z\ \chi_2\right)\biggr\}\\
+&\cos\left( \frac{\sqrt{3}}{2} \arctanh\  x\right)\times \biggl\{
\calc_2-\frac 16 \int_0^x dz
\frac{\sin\left( \frac{\sqrt{3}}{2} \arctanh\  z\right)}{z^3}\ \times\\
&\left(8z^2(z^2-1)\ \frac{d Z_\psi^0}{d z}
+8(z^2-1)\ \frac{d \c_2}{d z}-3z\ \chi_2\right)\biggr\}\,,
\end{split}
\eqlabel{gensolzc1}
\end{equation}
where $\calc_1$,  $\calc_2$ are integration constants.
For generic values of these constants we have asymptotically
\begin{equation}
\begin{split}
Z_\psi^1\sim &\sin\left( \frac{\sqrt{3}}{2} \arctanh\  x\right)
\left({\cal A}_\psi^s
 \left( 1+ \cdots \right) +   {\cal B}_\psi^s (1-x)^{3/4} + \cdots\right)\\
&+\cos\left( \frac{\sqrt{3}}{2} \arctanh\  x\right)
\left({\cal A}_\psi^c
 \left( 1+ \cdots \right) +   {\cal B}_\psi^c (1-x)^{3/4} + \cdots\right)
\end{split}
\eqlabel{zc1ass0}
\end{equation}
where ${\cal A}_\psi^s$,  ${\cal B}_\psi^s$, ${\cal A}_\psi^c$,  ${\cal B}_\psi^c$ 
are the ODE connection coefficients.
The integration constants  $\calc_1$,  $\calc_2$ should be chosen in
such a way that the boundary conditions for the matter fields
$ {\cal A}_\psi^s=0$, $ {\cal A}_\psi^c=0$ are satisfied.

To make
numerical analysis more convenient, we introduce functions $\calf_s(x)$,
$\calf_c(x)$ by
\begin{equation}
\begin{split}
Z_\psi^1(x)= (1-x^2)^{3/4}\left\{\sin\left( \frac{\sqrt{3}}{2}
\arctanh\  x\right)\ \calf_s(x)+\cos\left(
\frac{\sqrt{3}}{2} \arctanh\  x\right)\ \calf_c(x)\right\}\,.
\end{split}
\eqlabel{deffsfc}
\end{equation}
Redefining the radial coordinate
as in  Eq.~\eqref{redef} and using Eq.~\eqref{zc1} we find
that   $\calf_s(x)$,
$\calf_c(x)$ satisfy the equations
\begin{equation}
\begin{split}
&\frac{d \calf_s}{d y}-\frac{3}{4(1+y)}\ \calf_s = \\ &-\frac{\cos\left(
\frac{\sqrt{3}}{2} \arctanh\  y\right)}{y^{1/2}(1+y)^{3/2}}\
\left[\frac{3}{128}\ _2F_1\left(\frac 74,\frac 74; 3;
\frac{y}{1+y}\right)+g_\psi-\frac 43 (y+1)\ \frac{d g_\psi}{d y} \right]\,,
\end{split}
\eqlabel{fseq}
\end{equation}
\begin{equation}
\begin{split}
&\frac{d \calf_c}{d y} -\frac{3}{4(1+y)}\ \calf_c = \\ &\frac{\sin\left(
\frac{\sqrt{3}}{2} \arctanh\  y\right)}{y^{1/2}(1+y)^{3/2}}
\left[\frac{3}{128}\ _2F_1\left(\frac 74,\frac 74; 3; \frac{y}{1+y}\right)+g_\psi
-\frac 43 (y+1)\ \frac{d g_\psi}{d y} \right]\,.
\end{split}
\eqlabel{fceq}
\end{equation}
Asymptotics of the solutions $\calf_s(x)$,
$\calf_c(x)$ for  $y\to \infty$ are
\begin{equation}
\begin{split}
&\calf_s\to \calo(1)+\calo\left(y^{3/4}\right)\,,\\
&\calf_c\to \calo(1)+\calo\left(y^{3/4}\right)\,.
\end{split}
\eqlabel{leadas}
\end{equation}
Accordingly,
the initial conditions for
 the first-order ODEs \eqref{fseq} and \eqref{fceq} should be
 chosen in such a way that
the coefficients of the leading asymptotics in
\eqref{leadas} vanish. (This choice guarantees that the matter
fluctuations of the mode $Z_\psi^1$
do not change the fermionic mass parameter of the
dual gauge theory.)
Near the horizon,
  power series solutions to Eqs.~\eqref{fseq}, \eqref{fceq}
are
\begin{equation}
\calf_s=f_s^0+\left(-\frac 12 g_\psi^0-\frac{1}{32}\right)\ y^{1/2}+\frac 34 f_s^0\ y+\calo(y^{3/2})\,,
\eqlabel{fspss}
\end{equation}
\begin{equation}
\calf_c=f_c^0+\left(\frac{\sqrt{3}}{128}+\frac 34 f_c^0+\frac {\sqrt{3}}{8} g_\c^0\right)\ y
+\left(-\frac{3}{32} f_c^0+\frac{11\sqrt{3}}{8192}-\frac{37\sqrt{3}}{1536} g_\c^0\right)\ y^2+\calo(y^3)\,,
\eqlabel{fcpss}
\end{equation}
where  $f_s^0$, $f_c^0$ are integration constants, and
$g_\psi^0$ is chosen as in Eq.~\eqref{g0cfin}.
We use a ``shooting'' method to determine  $f_s^0$, $f_c^0$:
 these initial values should be tuned to ensure
that  $\calf_c$, $\calf_s$ remain finite in the limit
$y\to \infty$. We find
\begin{equation}
\begin{split}
f_s^0 \approx \,  &0.01964015(5)\,,\\
f_c^0 \approx \, &-0.01743333(5)\,,
\end{split}
\eqlabel{f0sc}
\end{equation}
where the brackets symbolize that there is an error in the corresponding digit.

\subsubsection{Step 4}
The solution to Eq.~\eqref{z21} regular at the horizon and obeying the
boundary condition \eqref{hb1} reads
\begin{equation}
\begin{split}
Z_2^1(x)=&\frac{2(x^2-1)}{3^{3/2}}\ \cali_{Z_2^1}^a(x)+\frac{4}{3^{3/2}}\
\cali_{Z_2^1}^b(x)
+\frac{2}{3^{3/2}}\ (x^2-1)\ln x \,\;  \cali_{Z_2^1}^c(x)\,,
\end{split}
\eqlabel{chihh}
\end{equation}
where the functions  $\cali_{Z_2^1}^a(x)$,  $\cali_{Z_2^1}^b(x)$,
 $\cali_{Z_2^1}^c(x)$ are given explicitly in Appendix \ref{appendix:structure_1}.
We verified that
\begin{equation}
\lim_{x\to 1_-}\cali_{Z_2^1}^a=\calo(1)\,, \;\;\;\;\;\;\;
   \lim_{x\to 1_-}\cali_{Z_2^1}^c=\calo(1)\,,
\eqlabel{chechz21ac}
\end{equation}
and thus the boundary condition \eqref{poledisp1}
translates into the equation
\begin{equation}
\lim_{x\to 1_-}\cali_{Z_2^1}^b=0\,.
\eqlabel{chechz21b}
\end{equation}
To determine the coefficient  $\b_2^\G$,
we solve Eq.~\eqref{chechz21b} numerically using the value of  $\b_2^v$ computed in
 \eqref{b1vnumer}.
We find
\begin{equation}
\b_2^\G\equiv \b_f^\Gamma \approx  0.9672(1)\,.
\eqlabel{b2gnumber}
\end{equation}

\subsection{Speed of sound and attenuation constant
to $\calo\left(\dd_1^2\right)$ in $\caln=2^* $ plasma}

We now turn to solving Eqs.~\eqref{za0}--\eqref{z11}
subject to boundary conditions \eqref{hb1}-\eqref{poledisp1}.
The computation is essentially
identical to the one in the previous Section,
thus we only highlight the main steps.

\subsubsection{Step 1}

Using the radial coordinate defined by Eq.~\eqref{redef}, we find that
the asymptotic behavior near the boundary of the general
 solution to Eq.~\eqref{za0} is given by
\begin{equation}
Z_\phi^0=
{\cal A}_\phi^0 y^{-1/2} + \cdots + {\cal B}_\phi^0 y^{-1/2} \log{y}
 + \cdots\,,
\eqlabel{oza0}
\end{equation}
where  ${\cal A}_\phi^0$, ${\cal B}_\phi^0$
 are the connection coefficients of the differential equation.
Introducing a new function $g_\phi(y)$ by
\begin{equation}
Z_\phi^0\equiv (1+y)^{-1/2}\ g_\phi(y)\,,
\eqlabel{defga}
\end{equation}
we obtain an inhomogeneous differential equation for $g_\phi$:
\begin{equation}
\begin{split}
&\frac{ d^2 g_\phi}{d y^2}
+ \frac{1}{y}\ \frac{d g_\phi}{ d y} -\frac{1}{4y(1+y)^2}\ g_\phi
-\frac{1}{96y(1+y)^2}\ _2F_1\left(\frac 32, \frac 32; 3;
 \frac{y}{1+y}\right) =0\,.
\end{split}
\eqlabel{gaeq}
\end{equation}
Near the horizon one can construct a series solution
parametrized by the integration constant  $g^0_\phi$
of the solution to the homogeneous equation
\begin{equation}
g_\phi=g^0_\phi+\left(\frac{1}{96}+\frac 14 g^0_\phi\right)\ y+
\left(-\frac{1}{384}-\frac{7}{64} g^0_\phi\right)\ y^2
+\calo(y^3)\,.
\eqlabel{ga1p}
\end{equation}
The integration constant $g^0_\phi$ is fixed by requiring that
\begin{equation}
g_\phi\sim \calo(1) \qquad {\rm as}\qquad y\to \infty\,.
\eqlabel{bbga}
\end{equation}
Using the ``shooting'' method we obtain
\begin{equation}
g_\phi^0 \approx  -0.08333333(3)\,.
\eqlabel{g0ares}
\end{equation}
We conjecture that the exact result is
\begin{equation}
g_\phi^0=-\frac{1}{12}\,.
\eqlabel{g0afin}
\end{equation}
(Note that a formal analytical solution
 to Eq.~\eqref{gaeq} is available. It allows to express
$g_\phi^0$ in terms of a certain  definite integral. Numerical 
evaluation of the integral confirms the result \eqref{g0afin}.)

\subsubsection{Step 2}
The  solution to Eq.~\eqref{z01} regular at the horizon and
satisfying the  boundary condition \eqref{hb1}
has the  form
\begin{equation}
Z_1^0(x)={8} (x^2-1)\ \cali_{Z_1^0}^a(x)-{8} \left[(x^2-1)\ln
x-2\right]\ \cali_{Z_1^0}^b(x)\,,
 \eqlabel{z10s}
\end{equation}
where
\begin{equation}
\begin{split}
\cali_{Z_1^0}^a(x)=&-\frac{x^2(3+x^2+2\ln x)}{4(1+x^2)^2}\ \beta_1^v
+\int_0^x\ dz\ \frac{(z^2-1)\ln z-2}{z(z^4-1)(1+z^2)^2}\ \times \\
&\biggl\{
24z^2
\left[-4z(z^2-1)\ \frac{d \alpha_1}{d z}
+(z^2+1)\ \alpha_1\right]\ Z_\phi^0\\
&+4(z^4-1)(z^2-1)^2\ \left(\frac{d\chi_2}{d z}\right)^2
+6z(z^2-1)^3\ \frac{d A_2}{d z} \biggr\}\,,
\end{split}
\eqlabel{z10a}
\end{equation}
\begin{equation}
\begin{split}
\cali_{Z_1^0}^b(x)=&-\frac{x^2}{2(1+x^2)^2}\ \beta_1^v
+\int_0^x\ dz\ \frac{1}{z(z^2+1)^3}\ \times \\
&\biggl\{
24z^2\left[-4z(z^2-1)\ \frac{d \alpha_1}{d z}
+(z^2+1)\ \alpha_1\right]\ Z_\phi^0\\
&+4(z^4-1)(z^2-1)^2\ \left(\frac{d \chi_2}{d z}
\right)^2+6z(z^2-1)^3\ \frac{d A_2}{d z} \biggr\}\,.
\end{split}
\eqlabel{z10b}
\end{equation}
We have verified that
\begin{equation}
\lim_{x\to 1_-} \cali_{Z_1^0}^a(x)= \calo(1)\,,
\eqlabel{checkz10a}
\end{equation}
and thus the boundary condition  \eqref{poledisp1} translates into
the equation
\begin{equation}
\lim_{x\to 1_-} \cali_{Z_1^0}^b(x)= 0\,.
\eqlabel{checkz10b}
\end{equation}
Solving Eq.~ \eqref{checkz20b} numerically for $\b_1^v$ we find
\begin{equation}
\b_1^v \approx -\frac {32}{\pi^2}\ \times\ 1.00000(1)\,,
\eqlabel{b2vnumer}
\end{equation}
where we factored out the value $- 32/\pi^2$ obtained
from thermodynamics (see Eq.~\eqref{cs}).

\subsubsection{Step 3}

Asymptotics of the general solution to Eq.~\eqref{za1} near the
boundary $y\to \infty$ is given by Eq.~\eqref{oza0} with different
coefficients  ${\cal A}_\phi^1$, ${\cal B}_\phi^1$. Rescaling the
dependent variable
\begin{equation}
Z_\phi^1= (1+y)^{-1/2}\ G_\phi(y)\,,
\eqlabel{defGa}
\end{equation}
we find that the function $G_\phi$
satisfies the following differential equation
\begin{equation}
\begin{split}
&\frac{d^2 G_\phi}{ d y^2}+ \frac{1}{y}\ \frac{d G_\phi}{ d y}
-\frac{1}{4y(1+y)^2}\ G_\phi
+\frac{\sqrt{3}}{96y(1+y)^2}\ _2F_1\left(\frac 32, \frac 32;
 3; \frac{y}{1+y}\right)\\
&-\frac{1}{2\sqrt{3}y (1+y)^2}\ \left[2(y+1)\ \frac{d g_\phi}{ d
y} -g_\phi\right]=0\,.
\end{split}
\eqlabel{Gaeq}
\end{equation}
One can construct a series solution regular near $y=0$
\begin{equation}
G_\phi=G_\phi^0 +\left(-\frac{\sqrt{3}}{144}-\frac{\sqrt{3}}{12}\
g_\phi^0 +\frac 14\ G_{\phi}^0\right)\ y+\calo(y^2)\,.
 \eqlabel{Ga1p}
\end{equation}
The integration constant $G^0_\phi$ is fixed by requiring that
\begin{equation}
G_\phi \sim \calo(1)\qquad {\rm as}\qquad y\to \infty\,.
\eqlabel{bbGa}
\end{equation}
 Using the ``shooting'' method and the value of
the constant  $g_\phi^0$ given by Eq.~\eqref{g0afin}
we find
\begin{equation}
G_\phi^0 \approx 0.059(0)\,.
\eqlabel{G0ares}
\end{equation}

\subsubsection{Step 4}
The solution to Eq.~\eqref{z11}
regular at the horizon and
obeying the boundary  condition \eqref{hb1} reads
\begin{equation}
\begin{split}
Z_1^1(x)=&\frac{2(x^2-1)}{3^{1/2}}\ \cali_{Z_1^1}^a(x)+\frac{4}{3^{1/2}}\ \cali_{Z_1^1}^b(x)
+\frac{2(x^2-1)\ln x}{3^{1/2}}\ \cali_{Z_1^1}^c(x)\,,
\end{split}
\eqlabel{alhh}
\end{equation}
where the functions $\cali_{Z_1^1}^a(x)$,  $\cali_{Z_1^1}^b(x)$,
 $\cali_{Z_1^1}^c(x)$ are given explicitly in Appendix
\ref{appendix:structure_2}.
We checked that
\begin{equation}
\lim_{x\to 1_-}\cali_{Z_1^1}^a\sim \calo(1)\,, \;\;\;\;\;\;\;
\lim_{x\to 1_-}\cali_{Z_1^1}^c \sim\calo(1)\,.
\eqlabel{chechz11ac}
\end{equation}
In view of Eq.~\eqref{chechz11ac},
the condition \eqref{poledisp1} becomes
\begin{equation}
\lim_{x\to 1_-}\cali_{Z_1^1}^b=0\,.
\eqlabel{chechz11b}
\end{equation}
To obtain the coefficient  $\b_1^\G$,
we solve  Eq.~\eqref{chechz11b} numerically using
 the value of $\b_1^v$ given by  Eq.~\eqref{b2vnumer}.
We find
\begin{equation}
\b_1^\G\equiv \b_b^\Gamma \approx 8.001(8)\,.
\eqlabel{b1gnumber}
\end{equation}
This completes out computation of the coefficients $\b_1^v$,
$\b_2^v$, $\b_1^\G$,  $\b_2^\G$: they are given, respectively, by
Eqs.~\eqref{b2vnumer}, \eqref{b1vnumer}, \eqref{b1gnumber},
 \eqref{b2gnumber}.

\section{Conclusion}
\label{sec_conclusions}

In this paper, we considered the problem of computing  the speed
of sound and the bulk viscosity of $\caln=2^*$ supersymmetric
$SU(N_c)$ gauge theory in the limit of large 't Hooft coupling and
large $N_c$, using the approach of gauge theory/gravity duality.
The computation can be done explicitly in the high temperature
regime, i.e. at a temperature much larger than the mass scale
$m_b$ and $m_f$ of the bosonic and fermionic components of the
chiral multiplets, where the metric of the dual gravitational
background is known.  Our results for the speed of sound and the
bulk viscosity computed in that regime are summarized in
Eqs.~\eqref{csa}, \eqref{csax}, \eqref{proporo}. It would be
interesting to extend the computation to the full parameter space
of the theory as well as to other theories with non-vanishing bulk
viscosity. It would also be interesting to compare our results
with a perturbative calculation of bulk viscosity in a
finite-temperature gauge theory at weak coupling.

\section*{Acknowledgments}
AB would like to thank O.~Aharony for valuable discussions. A.O.S.
would like to thank P.~Kovtun and D.~T.~Son for helpful
conversations, and L.~G.~Yaffe for correspondence. Research at
Perimeter Institute is supported in part by funds from NSERC of
Canada. AB gratefully   acknowledges  support by  NSERC Discovery
grant.

\appendix

\section{Energy density and pressure in $\caln=2^*$ gauge theory}
\label{holoren}

Energy density and pressure of $\caln=2^*$  SYM theory on the
boundary $\del\calm_5$ of a manifold $\calm_5$ with the metric
\eqref{ab} can be related to the renormalized stress-energy tensor
one-point function as follows
\begin{equation}
\ee=\sqrt{\sigma}N_\Sigma\ u^\mu u^\nu \langle T_{\mu\nu}\rangle
\,, \eqlabel{eden}
\end{equation}
\begin{equation}
P=\sqrt{\sigma}N_\Sigma\ \langle T_{x_1x_1}\rangle \ga^{x_1x_1}\,,
\eqlabel{pressure}
\end{equation}
where $u^\mu$ is the unit normal vector to a spacelike
hypersurface $\Sigma$ in $\del\calm_5$, $\sigma$ is the
determinant of the induced metric on $\Sigma$, and $N_\Sigma$ is
the norm of the timelike Killing vector in the metric \eqref{ab}.
The renormalized stress-energy tensor correlation functions are
determined from the boundary gravitational action (with the
appropriate counterterms added) in the procedure known as the
holographic renormalization. Holographic renormalization of
$\caln=2^*$ gauge theory on a constant curvature manifold was
studied in \cite{bh1}. Using the results for the renormalized
stress-energy tensor one-point functions \cite{bh1}, one finds
\begin{equation}
\cale=\frac{1}{96\pi G_5}
e^{4\xi}\left(18\beta-9\hat\r_{11}^2-12\hat\chi_0^2\hat\chi_{10}
+36\hat\r_{11}\hat\r_{10}-16\hat\chi_0^4\xi+36\hat\r_{11}^2\xi\right)\,,
\eqlabel{eneres}
\end{equation}
\begin{equation}
P=\frac{1}{96\pi G_5}
e^{4\xi}\left(6\beta+9\hat\r_{11}^2+12\hat\chi_0^2\hat\chi_{10}
-36\hat\r_{11}\hat\r_{10}+16\hat\chi_0^4\xi-36\hat\r_{11}^2\xi\right)\,,
\eqlabel{preres}
\end{equation}
where the parameters
$\beta,\xi,\rh_{10},\rh_{11},\chih_0,\chih_{10}$ are related to
physical masses and the temperature for generic values of $m_b/T$
and $m_f/T$  (see Section 6.4 of \cite{bl}).
In the limit $m_b/T\ll 1$ and $m_f/T \ll 1$, matching the high
temperature Pilch-Warner
 flow background \eqref{hightsol} with the general UV asymptotics
of the supergravity fields we find \cite{bl}
\begin{equation}
\begin{split}
&\beta=2,\qquad e^\xi=2^{1/2}\pi T\left(1-\frac{12}{\pi^2}\ \delta_1^2-\frac{1}{\pi}\ \delta_2^2\right)\,,\\
&\rh_{10}=\frac {4\ln2}{\pi}\ \dd_1,\qquad \rh_{11}=-\frac{8}{\pi}\ \dd_1\,,\\
&\chih_{0}=\frac{\sqrt{2\pi}}{\left[\Gamma\left(\ft
34\right)\right]^2}\ \dd_2,\qquad \chih_{10}=-\frac{2
\left[\Gamma\left(\ft 34\right)\right]^4}{\pi^2}\,.
\end{split}
\eqlabel{assyhight}
\end{equation}
Using Eq.~\eqref{assyhight},
 from Eqs.~\eqref{eneres}, \eqref{preres} we obtain
(to quadratic order in $\dd_1$,  $\dd_2$)
 the energy density and the pressure given in Eqs.~\eqref{finep}.

\section{Coefficients of Eq.~(4.17a)}
\label{appendix:coefficients_a}

\begin{equation}
\begin{split}
A_H (x) &=  3 \w^2 c_2' c_2^2
 c_1^2 \biggl(-c_2 c_1 q^2 c_1'-2 c_2' c_1^2 q^2+3 \w^2 c_2' c_2^2\biggr)\,, \\
 B_H (x) &= \w^2 c_2 c_1 \biggl(27 \w^2 c_2^2 c_1 c_2'^3-42 c_1^3 q^2 c_2'^3 +9 c_2^2 c_1 q^2 c_2' c_1'^2
-8 c_2^2 c_1^3 q^2 c_2'  \calp\\
&+8 c_2^3 c_1^2 q^2 c_1'  \calp
-3 c_2 c_1^2 q^2 c_2'^2 c_1'
+9 \w^2 c_2^3 c_2'^2  c_1'\biggr)\,,\\
 C_H (x) &=  \w^2 \biggl(-3 \w^2 c_2' c_2^3 q^2 c_1 c_1'+9 \w^4  c_2'^2 c_2^4
-8  q^2 c_1^4 c_2'^2 c_2^2 \calp+36 q^2 c_1^3 c_2'^3 c_1' c_2\\
&-24 q^2 c_1^4 c_2'^4
+16  q^2 c_1^3 c_2' c_1' c_2^3 \calp-8  q^2 c_1^2 c_1'^2 c_2^4 \calp+3
 c_2' c_1^3 q^4 c_1' c_2
+6  c_2'^2 c_1^4 q^4\\
&-12 q^2 c_1 c_1'^3 c_2^3 c_2'-15 \w^2  c_2'^2 c_1^2 q^2 c_2^2\biggr)\,,\\
 D_H (x) &=  16 q^2 c_1^2 \biggl(-c_1 c_2'+c_1' c_2\biggr) \biggl(24 c_1
 \w^2 (\a^{BG})' \calp c_2^3
+36 c_2^2 \w^2 c_2' (\a^{BG})' c_1'+3 c_2^2 c_1  \w^2 c_2' \frac{\del\calp}{\del\alpha}\\
&-c_2 q^2 c_1^2  \frac{\del\calp}{\del\alpha} c_1'-36 c_2 c_1 c_2'^2 \w^2 (\a^{BG})'
-24 c_2 q^2 c_1^3  \calp (\a^{BG})'
-2 q^2 c_1^3  c_2' \frac{\del\calp}{\del\alpha}\biggr)\,,\\
 E_H (x) &= 16 q^2 c_1^2 \biggl(-c_1 c_2'+c_1' c_2\biggr)
\biggl(8 c_1  \w^2 \c_{BG}' \calp c_2^3
+12 \w^2 c_2' \c_{BG}' c_1' c_2^2+3 c_2^2 c_1  \w^2 c_2' \frac{\del\calp}{\del\chi}\\
&-c_2 q^2 c_1^2 \frac{\del\calp}{\del\chi} c_1'-8 c_2 q^2 c_1^3
\calp \c_{BG}' -12 c_1 c_2'^2 \w^2 \c_{BG}' c_2-2 q^2 c_1^3  c_2'
\frac{\del\calp}{\del\chi}\biggr)\,. \nonumber
\end{split}
\eqlabel{eqsmainc}
\end{equation}
The prime denotes the derivative with respect to $r$.

\section{Coefficients of Eq.~(4.17b)}
\label{appendix:coefficients_b}

\begin{equation}
\begin{split}
A_\phi &= 12 c_2^2 c_1^2 c_2' \biggl(-c_2 c_1 q^2 c_1'-2 c_2'
c_1^2 q^2+ 3 \w^2 c_2' c_2^2\biggr)\,,\\
 B_\phi  &=  12  c_1 c_2' c_2 \biggl(-c_2 c_1 q^2 c_1'-2 c_2' c_1^2 q^2+3
\w^2 c_2' c_2^2\biggr)  \biggl(3 c_1  c_2'+ c_1' c_2\biggr)\,,\\
 C_\phi  &= -6 c_2^3 \w^2 c_2' c_1' q^2 c_1-192 c_2^6 \w^2 c_1^2
 (\a_{BG}')^2 \calp
-3 c_2^4 \w^2 c_2'^2 c_1^2 \frac{\del^2\calp}{\del\a^2}  \\
&- 30 c_2^2 \w^2 c_2'^2 q^2 c_1^2 +6 q^4 c_2' c_1^3 c_1' c_2+2 q^2
c_1^4 c_2^2 c_2'^2 \frac{\del^2\calp}{\del\a^2} -48 c_2^5 \w^2
c_2' \a_{BG}' c_1^2
 \frac{\del\calp}{\del\a} \\ &+
q^2 c_1^3 c_2^3 c_2' \frac{\del^2\calp}{\del\a^2} c_1' + 40 q^2
c_1^4 c_2' \a_{BG}' c_2^3 \frac{\del\calp}{\del\a}+8 q^2 c_1^3
\a_{BG}' c_2^4 \frac{\del\calp}{\del\a} c_1' +12 q^4 c_2'^2 c_1^4
\\ &+ 18 c_2'^2 c_2^4 \w^4 + 192 q^2 c_1^4 c_2^4
 (\a_{BG}')^2 \calp\,,\\
 D_\phi  &=
 - 2 c_1^2 c_2^2 \biggl(-c_2' c_2 c_1 \frac{\del^2\calp}{\del\a\del\c} c_1' q^2
+24 c_2' c_2^3
\a_{BG}' \frac{\del\calp}{\del\c} \w^2+3 c_2'^2 c_2^2
 \frac{\del^2\calp}{\del\a\del\c} \w^2
 \\ &- 2 c_2'^2 c_1^2 \frac{\del^2\calp}{\del\a\del\c} q^2 + 8
c_2' c_2^3 \w^2 \c_{BG}' \frac{\del\calp}{\del\a} +  64 c_2^4
\a_{BG}' \w^2 \c_{BG}' \calp  \\ &- 16 c_2' c_2 c_1^2 \a_{BG}'
\frac{\del\calp}{\del\c} q^2 - 64 c_2^2 c_1^2 \a_{BG}' q^2
\c_{BG}' \calp - 8 c_2' c_2 c_1^2 q^2 \c_{BG}'
\frac{\del\calp}{\del\a}  \\ &- 8 c_2^2 c_1 \a_{BG}'
\frac{\del\calp}{\del\c} c_1' q^2\biggr)\,,\\
 E_\phi  &= -  c_1^2 c_2^5 \w^2 \biggl(8 \a_{BG}' \calp c_2+
c_2' \frac{\del\calp}{\del\a}\biggr)\,,\\
 F_\phi  &=  c_1 c_2^4 \w^2 \biggl(8 \a_{BG}' \calp c_2+c_2' \frac{\del\calp}{\del\a}\biggr)
\biggl(-c_1 c_2'+c_1' c_2\biggr)\,.\nonumber
\end{split}
\end{equation}
The prime denotes the derivative with respect to $r$.

\section{Coefficients of Eq.~(4.17c)}
\label{appendix:coefficients_c}

\begin{equation}
\begin{split}
 A_\psi  &= 12 c_2^2 c_1^2 c_2'\biggl(-c_2 c_1 q^2 c_1'-2 c_2'
 c_1^2 q^2+3 \w^2 c_2' c_2^2\biggr)\,,\\
 B_\psi  &= 12 c_1 c_2' c_2\biggl(-c_2 c_1 q^2 c_1'-2 c_2' c_1^2 q^2+3 \w^2 c_2' c_2^2\biggr)
 \biggl(3 c_1 c_2' +  c_1' c_2\biggr)\,,\\
 C_\psi  &= 2 \biggl(3 q^2 c_1^3 c_2^3 c_2'
 \frac{\del^2\calp}{\del\c^2} c_1'+6 q^4 c_2' c_1^3 c_1' c_2
-30 c_2^2 \w^2 c_2'^2 q^2 c_1^2-64 c_2^6 \w^2 c_1^2 (\c_{BG}')^2 \calp  \\
&- 48 c_2^5 \w^2 c_2' \c_{BG}' c_1^2 \frac{\del\calp}{\del\c} -6
c_2^3 \w^2 c_2' c_1' q^2 c_1+6 q^2 c_1^4 c_2^2 (c_2')^2
\frac{\del^2\calp}{\del\c^2}+64 q^2 c_1^4 c_2^4 (\c_{BG}')^2 \calp
\\ &+8 q^2 c_1^3 \c_{BG}' c_2^4 \frac{\del\calp}{\del\c} c_1' +
40 q^2 c_1^4 c_2' \c_{BG}' c_2^3 \frac{\del\calp}{\del\c} -9 c_2^4
\w^2 c_2'^2 c_1^2 \frac{\del^2\calp}{\del\c^2}+ 18 c_2'^2 c_2^4
\w^4  \\ &+ 12 q^4 (c_2')^2 c_1^4\biggr)\,,\\
 D_\psi  &= -2  c_2^2 c_1^2 \biggl(-6 c_2'^2 c_1^2 \frac{\del^2\calp}{\del\a\del\c} q^2
+9 c_2'^2 c_2^2
\frac{\del^2\calp}{\del\a\del\c} \w^2-192 c_2^2 c_1^2 \a_{BG}'
 q^2 \c_{BG}' \calp \\ &+ 192 c_2^4 \a_{BG}'
\w^2 \c_{BG}' \calp - 8 c_2^2 c_1 \c_{BG}'
\frac{\del\calp}{\del\a} c_1' q^2-16 c_2' c_2 c_1^2 q^2 \c_{BG}'
\frac{\del\calp}{\del\a} \\ &- 3 c_2' c_2 c_1
\frac{\del^2\calp}{\del\a\del\c} c_1' q^2 +72 c_2' c_2^3 \a_{BG}'
\frac{\del\calp}{\del\c} \w^2 - 72 c_2' c_2 c_1^2 \a_{BG}'
\frac{\del\calp}{\del\c} q^2  \\ &+ 24 c_2' c_2^3 \w^2 \c_{BG}'
\frac{\del\calp}{\del\a}\biggr)\,,\\
 E_\psi  &=  -  c_2^5 c_1^2 \w^2 \biggl(3 \frac{\del\calp}{\del\c} c_2'
+8 c_2 \c_{BG}' \calp\biggr)\,,\\
 F_\psi  &= -  c_2^4 c_1 \w^2 \biggl(3 \frac{\del\calp}{\del\c} c_2'+8 c_2 \c_{BG}'
\calp\biggr) \biggl(c_1 c_2'-c_1' c_2\biggr)\,. \nonumber
\end{split}
\end{equation}
The prime denotes the derivative with respect to $r$.

\section{Structure functions of the solution \eqref{chihh}}
\label{appendix:structure_1}

\begin{equation}
\begin{split}
\cali_{Z_2^1}^a(x)=&\frac{3x^2(3+2\ln x+x^2)}{(1+x^2)^2}\ \b_2^\G+\biggl[
\frac{2x^2(x^2+3)^2}{(1+x^2)^3}\ \ln x-\ln(x^2+1)\\
&+\frac{4x^2(7+7x^2+2x^4)}{(1+x^2)^3}
\biggr]\ \b_2^v+\int_0^x\ dz \frac{(z^2-1)\ln z-2}{z(z^4-1)(1+z^2)^3}\ \times\\
&\biggl\{
24\sqrt{3} z^2(1+z^2)\left[-16z(z^2-1)\ \frac{d \chi_2}{d z}
+3(z^2+1)\ \chi_2\right]\ Z_\psi^1 (z)\\
&-3z(z^2-1)^3\ \frac{d Z_2^0}{d z}+6z^2(z^2-1)^2\ Z_2^0\\
&+48z^2\left[-8z(3z^2+1)(z^2-1)\
\frac{d \chi_2}{d z}+3(1+z^2)^2\ \chi_2\right]\  Z_\psi^0 (z)\\
&-24z(z^2+3)(z^2-1)^3\ \frac{d A_2}{d z} -16(1+z^2)^2(z^2-1)^3\
\left(\frac{d \c_2}{d z}\right)^2 \biggr\}\,,\nonumber
\end{split}
\end{equation}
\begin{equation}
\begin{split}
\cali_{Z_2^1}^b(x)=&\frac{6x^2}{(1+x^2)^2}\
 \b_2^\G+\frac{2x^2(9+6x^2+x^4)}{(1+x^2)^3}\ \b_2^v+\int_0^x dz \frac{1}{z(1+z^2)^4}
\times \\
&\biggl\{
24\sqrt{3} z^2(1+z^2)\left[
-16z(z^2-1)\ \frac{d \chi_2}{d z}+3(z^2+1)\ \chi_2\right]\ Z_\psi^1\\
&-3z(z^2-1)^3\ \frac{d Z_2^0}{d z}+6z^2(z^2-1)^2\ Z_2^0\\
&+48z^2\left[-8z(3z^2+1)(z^2-1)\
\frac{d \chi_2}{ d z}+3(1+z^2)^2\ \chi_2\right]\  Z_\psi^0\\
&-24z(z^2+3)(z^2-1)^3\ \frac{d A_2}{ d z} -16(1+z^2)^2(z^2-1)^3\
\left(\frac{d \c_2}{d z}\right)^2 \biggr\}\,,
\nonumber
\end{split}
\end{equation}
\begin{equation}
\begin{split}
\cali_{Z_2^1}^c(x)=&-\frac{6x^2}{(1+x^2)^2}\ \b_2^\G-\frac{2x^2(9+6x^2+x^4)}{(1+x^2)^3}\ 
\b_2^v-\int_0^x dz \frac{1}{z(1+z^2)^4}
\times \\
&\biggl\{
24\sqrt{3} z^2(1+z^2)\left[
-16z(z^2-1)\ \frac{d \chi_2}{d z}+3(z^2+1)\ \chi_2\right]\ Z_\psi^1\\
&-3z(z^2-1)^3\ \frac{d Z_2^0}{d z}+6z^2(z^2-1)^2\ Z_2^0\\
&+48z^2\left[-8z(3z^2+1)(z^2-1)\
\frac{d \chi_2}{ d z} +3(1+z^2)^2\ \chi_2\right]\  Z_\psi^0\\
&-24z(z^2+3)(z^2-1)^3\ \frac{d A_2}{d z} -16(1+z^2)^2(z^2-1)^3\
\left(\frac{d \c_2}{d z}\right)^2 \biggr\}\,. \nonumber
\end{split}
\end{equation}

\section{Structure functions of the solution \eqref{alhh}}
\label{appendix:structure_2}

\begin{equation}
\begin{split}
\cali_{Z_1^1}^a(x)=&\frac{x^2(3+2\ln x+x^2)}{(1+x^2)^2}\ \b_1^\G+\biggl(
\frac{2x^2(x^2+3)^2}{3(1+x^2)^3}\ \ln x-\frac 13 \ln(x^2+1)\\
&+\frac{4x^2(7+7x^2+2x^4)}{3(1+x^2)^3}
\biggr)\ \b_1^v+\int_0^x\ dz \frac{(z^2-1)\ln z-2}{z(z^4-1)(1+z^2)^3}\
\times\\
&\biggl\{
96\sqrt{3} z^2(1+z^2)\left[ -4z(z^2-1)\ \frac{d \a_1}{d z}
+(z^2+1)\ \a_1\right]\ Z_\phi^1\\
&-z(z^2-1)^3\ \frac{ d Z_1^0}{d z} +2z^2(z^2-1)^2\ Z_1^0\\
&+192z^2\left[-2z(3z^2+1)(z^2-1)\
\frac{d \a_1}{d z} +(1+z^2)^2\ \a_1\right]\  Z_\phi^0\\
&-8z(z^2+3)(z^2-1)^3\ \frac{d A_1}{d z} -16(1+z^2)^2(z^2-1)^3\
\left(\frac{d \a_1}{ d z}\right)^2 \biggr\}\,,\nonumber
\end{split}
\end{equation}
\begin{equation}
\begin{split}
\cali_{Z_1^1}^b(x)=&\frac{2x^2}{(1+x^2)^2}\ \b_1^\G
+\frac{2x^2(9+6x^2+x^4)}{3(1+x^2)^3}\ \b_1^v+\int_0^x dz \frac{1}{z(1+z^2)^4}
\times \\
&\biggl\{
96\sqrt{3} z^2(1+z^2)
\left[-4z(z^2-1)\ \frac{d \a_1}{d z}+(z^2+1)\ \a_1\right]\ Z_\phi^1\\
&-z(z^2-1)^3\ \frac{d Z_1^0}{d z} +2z^2(z^2-1)^2\ Z_1^0\\
&+192z^2\left[-2z(3z^2+1)(z^2-1)\
\frac{d \a_1}{d z} +(1+z^2)^2\ \a_1\right]\  Z_\phi^0\\
&-8z(z^2+3)(z^2-1)^3\ \frac{ d A_1}{d z} -16(1+z^2)^2(z^2-1)^3\
\left(\frac{d \a_1}{d z}\right)^2 \biggr\}\,,\nonumber
\end{split}
\end{equation}
\begin{equation}
\begin{split}
\cali_{Z_1^1}^c(x)=&-\frac{2x^2}{(1+x^2)^2}\ \b_1^\G
-\frac{2x^2(9+6x^2+x^4)}{3(1+x^2)^3}\ \b_1^v-\int_0^x dz \frac{1}{z(1+z^2)^4}
\times \\
&\biggl\{
96\sqrt{3} z^2(1+z^2)\left[-4z(z^2-1)\ \frac{d \a_1}{d z}+(z^2+1)\
 \a_1\right]\ Z_\phi^1\\
&-z(z^2-1)^3\ \frac{ d Z_1^0}{d z} +2z^2(z^2-1)^2\ Z_1^0\\
&+192z^2\left[ -2z(3z^2+1)(z^2-1)\
\frac{d \a_1}{d z} +(1+z^2)^2\ \a_1\right]\  Z_\phi^0\\
&-8z(z^2+3)(z^2-1)^3\ \frac{d A_1}{ d z} -16(1+z^2)^2(z^2-1)^3\
\left(\frac{d \a_1}{d z}\right)^2 \biggr\}\,.\nonumber
\end{split}
\end{equation}

\end{document}